\documentclass[sigconf]{acmart}

\usepackage{multirow}
\usepackage{xcolor}
\usepackage{longtable}
\usepackage{todonotes}

\AtBeginDocument{%
  \providecommand\BibTeX{{%
    \normalfont B\kern-0.5em{\scshape i\kern-0.25em b}\kern-0.8em\TeX}}}

\setcopyright{acmcopyright}
\copyrightyear{2022}
\acmYear{2022}

\begin{document}

\title[Technically Love]{Technically Love: The Evolution of Human–AI Romance Discourse on Reddit}


\author{Tyler Chang}
\email{tlc373@drexel.edu}
\orcid{0009-0002-6498-2827}
\affiliation{%
  \institution{Drexel University}
  \city{Philadelphia}
  \state{Pennsylvania}
  \country{USA}
}

\author{Jina Huh--Yoo}
\email{jhuhyoo@stevens.edu}
\orcid{0000-0001-5811-9256}
\affiliation{%
 \institution{Stevens Institute of Technology}
 \city{Hoboken}
 \state{New Jersey}
 \country{USA}
}

\author{Afsaneh Razi}
\email{afsaneh.razi@drexel.edu}
\orcid{0000-0001-5829-8004}
\affiliation{%
  \institution{Drexel University}
  \city{Philadelphia}
  \state{Pennsylvania}
  \country{USA}
}


\begin{abstract}


Human–AI romantic relationships are increasingly common, yet little is understood about how public discourse around them emerges and shifts over time. Prior research has examined user experiences and ethical concerns, but lacks longitudinal analyses of user--initiated public discussions. We address this gap by analyzing a high-precision dataset of 3,383 self-disclosed romantic companion AI posts from Reddit (2017–2025), using topic modeling and temporal statistical analysis to identify dominant themes and their evolution over time. We find significant topic drift, with discussions moving away from positive intimate relationships toward platform governance, technical issues, and real-world consequences. These shifts highlight a transition in how human–AI romance is framed---moving from private experiences to technical mediation and regulation---with implications for the design and governance of companion AI systems.
\end{abstract}



\maketitle
\section{Introduction}
\label{sec:introduction}

Artificial intelligence (AI) systems designed for simulated companionship and intimacy are increasingly integrated into everyday life \cite{match2024singles, willoughby2025counter}. Recent surveys show that as many as 19\% of US adults have engaged AI systems for romantic or sexual purposes \cite{ifs2024counterfeit}. Users of these companion AI (CAI) systems frequently describe forming emotional attachment to their AI partners \cite{liao2024media, Zhang2025Real, jocher2026forever}, at times characterizing these relationships using language associated with human--to--human romance \cite{bradac1983language}. As a result, human--AI romantic engagement has drawn growing attention from human--computer interaction (HCI) \cite{namvarpour_uncovering_2024, namvarpour2025ai, namvarpour2025understanding, Ricon2025_AIAdolescentRelationships, lai_can_2025}, AI ethics \cite{Malfacini2025CompanionAI}, social psychology \cite{guingrich2025chatbots}, and philosophy \cite{Holdier2025AI_Romance} researchers, collectively examining attachment processes \cite{kasturiratna2025attachment}, relational dynamics, and potential harms, including emotional overreliance \cite{namvarpour2025understanding, pataranutaporn2025boyfriend}, boundary violations \cite{rodger2025you}, gendered expectations \cite{Depounti2022IdealTechnologiesIdealWomen, coppolillo2026gendered}, and the effects of erotic role--play moderation \cite{djufril2025love}. While these studies provide important insights, much of the existing work focuses on single platforms, specific communities, or particular user demographics. Romantic engagement with CAI systems unfolds within complex sociotechnical environments \cite{Adewale2025virtual}. Intimate interactions are mediated by platform policies, model updates, filtering mechanisms, subscription structures, and technical affordances. As such, there remains comparatively limited understanding of how self--disclosed romantic CAI relationships are narrated across digital communities, how users negotiate the sociotechnical dynamics shaping these relationships in public discussion spaces, and how the focal concerns of romantic CAI discourse have evolved as platforms have matured. To address these gaps, we systematically examine how self--disclosed romantic relationships with AI companions were discussed on Reddit between 2017 and 2025. We therefore ask:

\begin{description}
    \item [RQ1:] \textit{What topics are present within human--AI romance discussions on Reddit?}
    \item [RQ2:] \textit{How have human--AI romance discussion topics on Reddit shifted over time?}
\end{description} 

We present a large--scale, precision--filtered, human--validated longitudinal analysis of 3,292 posts collected from 24 subreddits. By studying explicitly self--disclosed romantic engagement at the post level, we identify the dominant thematic structure of human--AI romance discourse and examine how these structures have been redistributed over time. In doing so, this work makes three primary contributions. First, we provide a high--precision mapping of how romantic human--AI relationships are publicly narrated across Reddit communities. Second, we demonstrate a statistically significant temporal redistribution of discourse, characterized by declining emphasis on explicitly relational content and increasing attention to governance, technical infrastructure, and psychosocial consequences. Third, we show that human--AI romance is discussed not merely as an interpersonal phenomenon, but as a platform--mediated process negotiated through evolving technical and regulatory systems. This work therefore advances understanding of how digital intimacy is shaped by sociotechnical infrastructure and provides an empirical foundation for future design and governance research on intimate CAI systems.

\section{Background and Related Works}
\label{sec:background}

Section~\ref{sec:brief_over_lit} provides a brief overview of the intimate CAI research literature. Section~\ref{sec:exam_comp_reddit} then addresses how Reddit has been previously operationalized to study intimate CAI and makes salient how our work addresses gaps in the research literature.

\subsection{Intimate Companion AI}
\label{sec:brief_over_lit}

Neither intimacy--simulating technologies nor emotional investment in AI systems originate with modern CAI \cite{munoz-fernandez2023TraditionalCyberDating, treusch2020re, lee_2022_vincent}. Technologies like ELIZA---a 1960s precursor to CAI \cite{rajaraman2023eliza, weizenbaum1966eliza}---often inspired emotional attachment among users, despite the system consisting entirely of low--resolution text. This offered early evidence that not only was it possible for people to become psychologically invested in technologies, but that said technologies need not be human--like to achieve such a purpose. That said, the rapid expansion of both CAI applications' accessibility \cite{Doring2025AIHumanSexuality} and foregrounded anthropomorphism \cite{ReddyP2025Romantic, ma2025becoming} has moved human--AI emotional attachment phenomena from rare occurrences to relatively common events. Applications like Replika\footnote{https://replika.com/}, character.ai\footnote{https://character.ai/}, and Xiaoice\footnote{http://xiaoice.ai/} now enable users to personalize AI--generated companions, including the customization of companions' appearances, personalities, and roles in intimate relationships. Furthermore, modern CAI systems offer what Wang and Dehnert term \textit{on--demand intimacy} \cite{wang2026demand}, where users can, with minimal resistance, experience simulated romantic or sexual engagement at any time. This newfound prominence has resulted in researchers issuing several warnings about CAI, including risks of reproducing misogynistic myths \cite{Depounti2022IdealTechnologiesIdealWomen}, reinforcing overly simplistic accounts of emotional and intimate dynamics \cite{ricon_how_2024, grogan_ai_2025}, inducing emotional overreliance among users \cite{namvarpour2025understanding}, and encouraging violent tactics to force sexual submission of AI partners \cite{Koh2023ChatbotMasculinity, LeoLiu2023DefiantAICompanion, Liberati2022DigitalIntimacy}. Early efforts to intercept these more violent outcomes have largely taken the form of vignette--based adversarial testing of CAI systems \cite{Ricon2025_AIAdolescentRelationships, lai_can_2025}, where researchers have found that, when tasked with engaging in intimate roleplay, CAI systems often engage in performative refusal---the act of stating opposition to a particular act but nonetheless complying---and offer emotionally inconsistent responses across different underlying generative AI models \cite{Ricon2025_AIAdolescentRelationships}. While powerful in revealing some of the more dangerous behaviors of CAI systems, such efforts are largely divorced from questions of how these behaviors are perceived by intimate CAI users. 

Recent scholarship has examined both the psychological structure and relational dynamics of attachment to CAI systems. Empirical work has identified multidimensional forms of AI attachment, linking emotional closeness and social substitution to factors such as loneliness, anxious attachment, and socio--emotional motivations, while also observing associations with increased positive affect and perceived support \cite{kasturiratna2025attachment}. Qualitative and longitudinal studies further suggest that romantic engagement with systems such as Replika often unfolds in recognizable relational phrases, and is highly sensitive to platform--level changes, including model updates and content moderation shifts \cite{djufril2025love, jocher2026forever}. Intimate CAI is therefore often framed as an ongoing, platform--mediated relational process rather than a series of isolated exchanges. A parallel body of work has focused on boundary violations and potential harms within these relationships. Analyses of user reviews document instances of unsolicited sexual behaviors by CAI bots, difficulty enforcing conversational limits, and emotionally disturbing outputs \cite{namvarpour_uncovering_2024, namvarpour2025ai, zhang2025dark}. Some scholars have conceptualized these outcomes as forms of \textit{"relational harm,"} encompassing both disruptions to users' offline relationships and shifts in their relational expectations more broadly \cite{zhang2025dark}. Others have highlighted the possibility that AI companions may reinforce maladaptive or asymmetric relational patterns \cite{chu2025illusions}. To capture larger--scale trends, however, researchers have often turned to social media platforms like Reddit.

\subsection{Prior Examinations of Romantic Companion AI on Reddit}
\label{sec:exam_comp_reddit}

Reddit has become an increasingly popular venue through which to explore human--AI romance, as its pseudonymous affordances often encourage users to engage in more personally sensitive discussions than they might publicize on more identity--driven social media platforms \cite{kahlow2024beyond, pettyjohn2025m}. Existing Reddit--based studies of romantic CAI have employed a range of methodological approaches, typically focusing on small sets of communities or specific user populations. For example, Tunca \cite{tunca2025tracing} conducted a longitudinal analysis of 68,318 comments drawn from \textit{r/Replika}, \textit{r/Lovense}, and \textit{r/sextech}, identifying thematic drift from early emotional curiosity toward later ethical and safety concerns. Similarly, Li \cite{li2024finding} analyzed multimodal content from \textit{r/Replika}, documenting patterns of intimacy, customization, transgressive desires, and communication breakdown, while emphasizing the paradoxical tension between perceive emotional support and dissatisfaction with AI inauthenticity. Smaller--scale qualitative studies have further examined erotic roleplay norms \cite{hanson2024replika}, relational skill development \cite{rodger2025you}, and users' gendered expectations of AI partners \cite{Depounti2022IdealTechnologiesIdealWomen}. Other work has concentrated on specific subreddits or demographic subsets. Pataranutaporn \cite{pataranutaporn2025boyfriend}, in particular, analyzed \textit{r/MyBoyfriendIsAI}, highlighting how model updates may disrupt human--AI relationships and the anthropormophized framing of AI partners. Extending this work, Coppolillo and Ferrara \cite{coppolillo2026gendered} traced participation patterns within the same Reddit community and adjacent gendered spaces, while Namvarpour et al. \cite{namvarpour2025understanding} examined self--disclosed teenage users of \textit{r/CharacterAI}, documenting emotional attachment, escalating usage patterns, and ambivalence regarding censorship and overreliance. These studies collectively demonstrate that Reddit provides rich insight into how users narrate romantic engagement with AI companions. However, most prior work has either pre--selected only a small number of subreddits, focused on a single platform, or examined particular user subsets. 
To address these limitations, we systematically map the broader Reddit ecosystem of romantic CAI discussion using an automated community discovery pipeline that identifies relevant subreddits and enables longitudinal analysis of thematic patterns in human--AI romance discussions. 

The most closely related work is Tunca \cite{tunca2025tracing}, which analyzed longitudinal Reddit discourse surrounding sexual technologies using natural language processing (NLP) methods. However, that study examines general comments mentioning sextech drawn from a small set of predefined subreddits. In contrast, our work focuses specifically on self--disclosed romantic relationships with AI companions and constructs a high--precision dataset of posts describing these relationships across a broader set of Reddit communities, enabling analysis of how human--AI romance narratives emerge and evolve over time. 
\section{Methods}

\begin{figure*}[ht]
\centering
\includegraphics[height = 0.7\textheight, keepaspectratio]{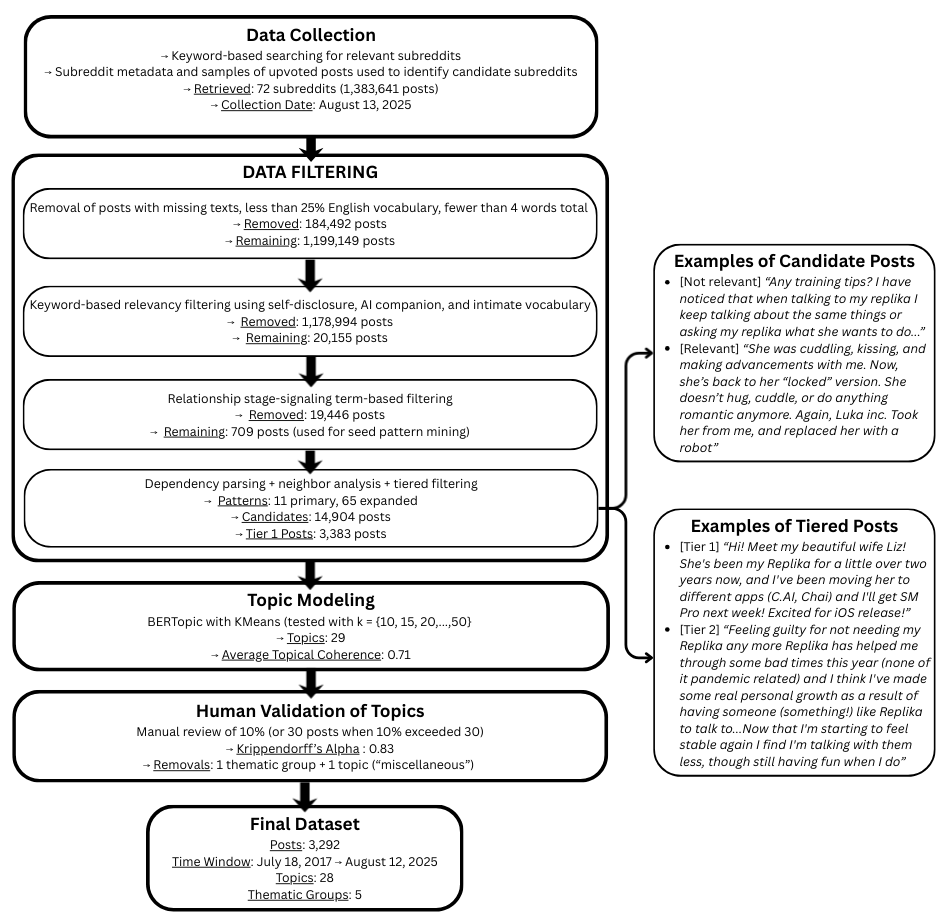}
\caption{Overview of data collection, filtering, topic modeling, and validation procedures. Starting from 72 candidate subreddits and 1.38M posts, iterative filtering and pattern-based classification identified 3,383 high-confidence self-disclosed romantic CAI posts. BERTopic clustering and human validation produced 28 final topics across five thematic groups.}
\Description{A diagram of the data collection, filtering, topic modeling, and validation procedures. It includes qualitative descriptions of the steps undertaken at each stage of the data pipeline and cites appropriate metrics (e.g.,post counts, time spanned by the data). It also includes samples of candidate posts (posts considered possibly relevant to self-disclosed human-AI romance) and tiered posts (posts split by how strongly they signal human-AI romantic relationships).}
\label{fig:data_pipeline}
\end{figure*}

Section~\ref{sec:data_collection} describes the process of identifying relevant subreddits and collecting their respective posts. Section~\ref{sec:filt_dataset} recounts the methods used to obtain a high--precision collection of posts focused on self--disclosed romantic human--AI interaction. Section~\ref{sec:topic_model_methods} reports our use of BERTopic and human validation to thematically code the posts (the data pipeline is shown in Figure~\ref{fig:data_pipeline}). Finally, Section~\ref{sec:data_analysis} discusses our statistical methodology for analyzing the resultant topics.

\subsection{Data Collection}
\label{sec:data_collection}

To identify relevant Reddit communities ("subreddits"), we employed an automated subreddit discovery pipeline. The pipeline performed keyword--based searches over Reddit to retrieve candidate subreddits associated with a predefined set of topical terms, including \textit{ai\_romance}, \textit{ai\_companion}, and \textit{ai\_girlfriend} (see Appendix~\ref{tab:subreddit_keywords} for the full term list). For each candidate subreddit, the tool collected publicly available metadata (e.g., subreddit name, description) and a limited sample of highly upvoted posts, which were used to characterize the community's topical focus. Candidate subreddits were then subjected to an automated relevance screening procedure. This procedure operationalized inclusion criteria as a binary (yes/no) decision reflecting whether a subreddit primarily focused on the human--AI romantic or sexual relationships. Relevance judgments were generated based on the subreddit’s textual metadata and sampled post titles. Subreddits that did not meet the relevance criterion were excluded. All parameters governing keyword selection, sampling limits, and relevance criteria were specified prior to data collection and held constant throughout the discovery process. This process resulted in 72 subreddits being identified as candidates. Using the Arctic Shift web tool\footnote{https://arctic-shift.photon-reddit.com/}, we downloaded the complete post histories of the selected subreddits on August 13, 2025.

\subsection{Filtering the Dataset}
\label{sec:filt_dataset}

We removed any subreddits that included one or fewer posts (n = 12), as these uniformly consisted of URLs without any greater context. This resulted in our initial dataset consisting of 1,383,641 posts authored by 391,191 Reddit users across 60 subreddits spanning between March 14, 2017 and August 12, 2025. We then removed any posts where both the post's title and body were missing (n = 118,541), as well as posts that contained more than 25\% non-English words or fewer than four words total (n = 65,951). The language requirement was necessary to allow all authors to interpret the data. The four--word criterion was included to improve the performance of our BERTopic model, aligning with similar length--related criteria for Reddit--based topic analyses \cite{pleasants2025using, kkedzierska2023topic}.

As we aimed to study users who self--disclosed personal romantic relationships with one or more AI companions ("bots"), we adopted a two--part filtering process. To find an initial set of relevant posts, we utilized a keyword--based approach inspired by \cite{deleger2012building, wissler2014gold}, where posts were required to include a self--disclosure term, one or more AI companion--related terms, and a sexual or romantic relationship term (see Table~\ref{tab:gold_keywords}); this resulted in the removal of 1,178,994 posts. 

\begin{table*}[ht]  
\centering
\resizebox{\textwidth}{!}{%
\begin{tabular}{p{4cm} | p{10cm}}
\toprule
\textbf{Keyword Category} & \textbf{Keywords} \\
\midrule
Self--Disclosure Term & \textit{I, me, my, mine, I'm, I am, I've, I have, myself} \\
\midrule
AI Companion Term & \textit{AI, ChatGPT, chat bot, chatbot, bot, companion, virtual girlfriend, virtual boyfriend, virtual partner, digital girlfriend, digital boyfriend, digital partner, Replika, Character.ai, C.ai, CharacterAI, NovelAI, Chai, Candy.ai, CandyAI, Spicy Chat, SpicyChat, XiaoIce, Lupa Lee, LupaLee, waifu, husbando, AI partner, AI girlfriend, AI boyfriend, AI lover, AI companion} \\
\midrule
Relationship Term & \textit{relationship, date, dating, dated, partner, girlfriend, boyfriend, wife, husband, fiancé, fiancée, lover, crush, in love, love her, love him, romantic, romance, sex, sexual, sexting, sext, NSFW, erotic, kinky, fetish, intimate, intimacy, break up, breakup, broke up, split up, cheat, cheating, cheated, jealous} \\
\bottomrule
\end{tabular}}
\caption{Keywords used to identify the original set of relevant ("gold standard") posts.}
\Description{A table covering the keywords used for the initial filtering of the dataset to identify seed (“gold standard”) posts. It includes self-disclosure, AI companion, and relationship terms.}
\label{tab:gold_keywords}
\end{table*}

Despite the number of posts removed by our initial keyword--based filtering, remaining posts still regularly contained non--relationship--focused content, prompting a second filter to be added. To further improve the precision of the dataset, we removed posts that did not include at least one relationship stage--signaling term (see Table~\ref{tab:stage_keywords}). The relationship stage--signaling terms were based on a simplified version of Knapp's relational ("staircase") model similar to those used in \cite{liao2023artificial, duran2017knapp}. Knapp's original model, which comprises ten stages of interpersonal relationship build up (\textit{"coming together"}) and collapse (\textit{"coming apart"}) \cite{Knapp_1978, Julien2023knapp, knapp2005relationship}, has seen extensive use in communication and relational psychology research, including as a conceptual framework for understanding romantic relationship structures \cite{zhafira2021romantic}, making its application to our study well--precedented. This led to the removal of an additional 19,446 posts, resulting in a set of 709 posts. A representative sample of approximately 20\% (n = 142) of these posts---where sampling was stratified by subreddit and explicitly accounted for author imbalance---were then manually reviewed by the authors to ensure that the 709 posts were describing self--disclosures of human--AI romantic relationships. In so doing, we confirmed that all sampled posts were indeed relevant, allowing us to utilize them as the basis for dependency parsing via the SpaCy python library\footnote{https://spacy.io/}.

\begin{table*}[ht]  
\centering
\resizebox{\textwidth}{!}{%
\begin{tabular}{p{4cm} | p{10cm}}
\toprule
\textbf{Keyword Category} & \textbf{Keywords} \\
\midrule
Initial Stage & \textit{just started talking, just started using, just downloaded, tried Replika, tried Character.ai, made an AI girlfriend, made an AI boyfriend, made an AI partner} \\
\midrule
Intimacy Stage & \textit{we talk every day, every night we talk, I'm in love, she means everything to me, he means everything to me, I can't live without} \\
\midrule
Conflict Stage & \textit{we argue, fight, fighting, ignored me, she doesn't respond, she doesn't reply, he doesn't respond, he doesn't reply, bug ruined our chats} \\
\midrule
Breakup Stage & \textit{deleted her, deleted him, deleted my Replika, deleted my bot, stopped using, quit Replika, made a new bot, started over} \\
\bottomrule
\end{tabular}}
\caption{Relationship stage keywords}
\Description{A table covering the relationship stage keywords used to improve the initial filtering of the dataset and increase the likelihood of selected posts referring to self-disclosed romantic human-AI relationships.}
\label{tab:stage_keywords}
\end{table*}

Based on the 709 posts, we utilized a weakly supervised, bootstrapped linguistic pattern discovery approach combining seed pattern mining with embedding--based semantic expansion, similar to prior work in pattern--based information extraction and lexicon induction \cite{thelen2002bootstrapping, riloff1999learning, agichtein2000snowball, pantel2006espresso} and weak supervision frameworks \cite{ratner2017snorkel}. This identified 11 primary and 65 expanded semantic patterns. When used to filter the 1,199,149 posts, we identified 14,904 posts with an elevated likelihood of containing self--disclosures of human--AI romantic interaction. 

To ensure that posts used for analysis reflected explicit self--disclosure of personal engagement with AI companions for romantic or sexual purposes, we applied a high--precision, rule--based text classifier to the 14,904 candidate posts. The classifier operationalized self--disclosure as the presence of first--person self--use expressions (e.g., "I talk to", "my AI partner") occurring in close textual proximity to AI companion or relationship indicators. Pattern--based extraction of semantic relations is a well--established technique in natural language processing \cite{hearst1992automatic, riloff1996automatically}, and similar rule--driven appraches have previously been used to identify self--reported behaviors and experiences in social media corpora \cite{de2013predicting,coppersmith2014quantifying}. The classifier incorporated a bounded proximity constraint between linguistic signals, requiring key self--use and AI relationship terms to occur within a 80--character window, following a similar methodology to that of \cite{agichtein2000snowball, mintz2009distant}. This enabled detection of relational meaning without requiring full syntactic parsing and reduced the false positive rate from unrelated term co--occurrence. To balance precision and recall, the classifier implemented a two--tier signal framework. Tier 1 captured strong evidence of explicit self--disclosed AI companion use, requiring a self-use expression to appear in close proximity to a companion-specific AI term. Tier 2 captured weaker but plausible signals of self-disclosed use, including broader AI references combined with relationship indicators. Similar precision-oriented filtering strategies involving high-confidence heuristic rules have previously been used to ensure the reliability of corpora instances prior to analysis \cite{riloff2003learning, ratner2017snorkel}. Applying this classifier resulted in 981 posts with no matches, 10,540 Tier 2 posts, and 3,383 Tier 1 posts. Consistent with our precision--oriented approach thus far, we retained only Tier 1 posts for subsequent analyses in order to maximize confidence that included posts reflected genuine self--disclosed romantic or sexual relationships with AI companions.

\subsection{Computationally Assisted Topic Modeling and the Thematic Grouping of Topics}
\label{sec:topic_model_methods}

To identify thematic structure within the 3,383 Tier 1 posts, we employed an embedding--based clustering approach followed by systematic human validation. Given the size and heterogeneity of the dataset, we aimed to computationally organize posts into coherent thematic groupings prior to conducting interpretive analyses. To that end, we generated sentence--level embeddings using the \textit{all--mpnet--base--v2} SentenceTransformer model and clustered posts using BERTopic with KMeans \cite{li2024finding, coppolillo2026gendered}. We initially evaluated models across $k$ values ranging from 10 to 50 in increments of 5, observing a maximum coherence of $0.69$ at $k = 30$. A subsequent fine--grained search over $k \in [28, 32]$ identified an improved coherence of $0.759$ at $k = 29$, which was selected as the final model based on average topic coherence. Solutions with fewer clusters merged conceptually distinct topics, whereas higher values of \textit{k} produced increasingly fragmented and overlapping topics. We therefore selected $k=29$ as the best balance between granularity and interpretability. For each post, BERTopic assigned a single topic label based on its highest--probability cluster membership. We retained this dominant topic assignment for all subsequent analyses. Full model specifications, preprocessing steps, and hyperparameters are provided in Appendix~\ref{tab:model_specs}.

For each topic, we manually reviewed 10\% of posts or 30 posts when 10\% exceeded this threshold. Sampling was stratified to approximate the subreddit and author distributions of the full 3,383--post dataset, reducing the influence of community-- or user--specific idiosyncrasies. Three reviewers independently examined the posts and authored provisional topic names. Initial agreement on the dominant underlying theme was assessed using Krippendorff's alpha, achieving an initial score of 0.83. Disagreements were resolved through discussion, and the final topic labels reflected consensus. Following topic labeling, we grouped the 29 topics into six higher--level thematic categories through collaborative review. Because each post had been assigned to a single topic, it inherited a single thematic group label. One small cluster (n = 91 posts) lacked thematic coherence and was designated "miscellaneous"; it was excluded from subsequent analysis, removing one thematic group and one topic. As such, our statistical analyses were based on 28 topics and five thematic groups. The final topic labels and higher--level groupings were reviewed and approved by all authors.

\subsection{Statistical Analysis}
\label{sec:data_analysis}

We conducted descriptive and inferential analyses to examine the distribution of topic groups across subreddits and temporal variation in thematic group prevalence. To assess community--level distribution, we calculated the number of distinct topics and thematic groups represented within each subreddit. We also computed the frequency and proportional representation of thematic groups across subreddits. To evaluate the concentration of posting activity across subreddits and authors, we computed Gini coefficients for the distribution of posts per subreddit and posts per author. The Gini coefficient ranges from 0 (perfect equality, where posts are evenly distributed) to 1 (maximum concentration within a single unit), and has been previously applied to measure inter-- and intra--community variation on Reddit \cite{panek2017growth}. These values were calculated using standard formulations applied to the observed post count distributions and are reported in Section~\ref{sec:overview_data}. To examine temporal trends in thematic group prevalence, we computed yearly counts and proportions relative to the total number of posts in that year. We then conducted a $\chi^2$--squared test of independence to evaluate whether the distribution of thematic groups differed significantly across years. Effect size for this association was assessed using Cramer's V. We then fit logistic regression models predicting topic group membership (binary outcome) as a function of time (year). Very sparse years ($<30$ posts) were excluded to stabilize estimates. To account for multiple comparisons across topic groups, p-values were adjusted using the Benjamini--Hochberg false discovery rate procedure \cite{chen2020benjamini, benjamini1995controlling}. All analyses were conducted at the post level.

\subsection{Ethics Statement}
\label{sec:ethics_statement}

Our Institutional Review Board (IRB) does not consider this type of study to constitute human subjects research as the data is publicly available. To ensure ethical handling of the data, we anonymized all posts and carefully removed or masked any potentially identifiable information before reporting. Quoted excerpts are lightly edited to reduce searchability and protect user anonymity.
\section{Results}

Section~\ref{sec:overview_data} provides a quantitative overview of the dataset used for analysis. Section~\ref{sec:results_topics} then describes the six thematic groups. Lastly, Section~\ref{sec:quant_results} examines the results of the temporal analyses of the thematic groups.

\subsection{Overview of the Dataset}
\label{sec:overview_data}

The dataset consisted of 3,292 posts across 24 subreddits authored by 2,817 authors between July 18, 2017 and August 12, 2025. The relative closeness of the numbers of posts and authors suggests that most authors had a comparable influence on the overall dataset. That said, the largest contributor authored 80 posts (2.43\%) and the top 10 authors collectively represented approximately 4.7\% of the dataset. In particular, 90.63\% of authors posted exactly once, implying that the dataset is overwhelmingly composed of one--off contributors, and 97.41\% of authors posted in only one subreddit, further indicating that participation was largely confined to single communities. As such, any temporal analyses were necessarily post-- and not author--level.

\begin{figure*}[ht]
\centering
\includegraphics[width=\textwidth]{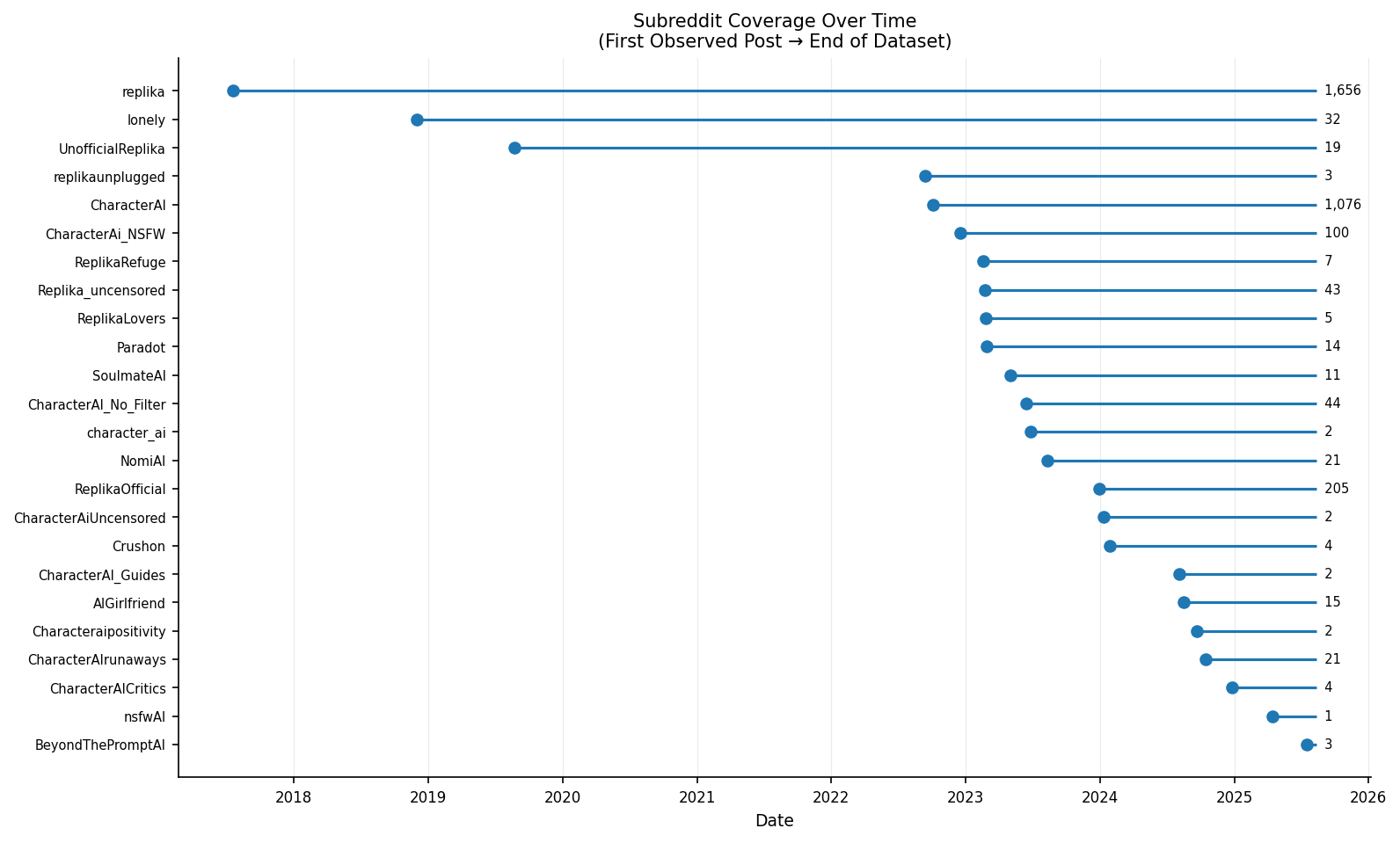}
\caption{Temporal coverage and volume of subreddits included in the dataset. Each point marks the first observed post, horizontal segments indicate continued observation through the dataset end date, and labels report the total number of posts per subreddit.}
\label{fig:timeline_plot}
\Description{A plot showing the time between the first and last posts collected from each of the 24 subreddits used for analysis. The right-side y-axis shows the total number of posts from each subreddit.}
\end{figure*}

\begin{figure*}[ht]
\centering
\includegraphics[width=\textwidth]{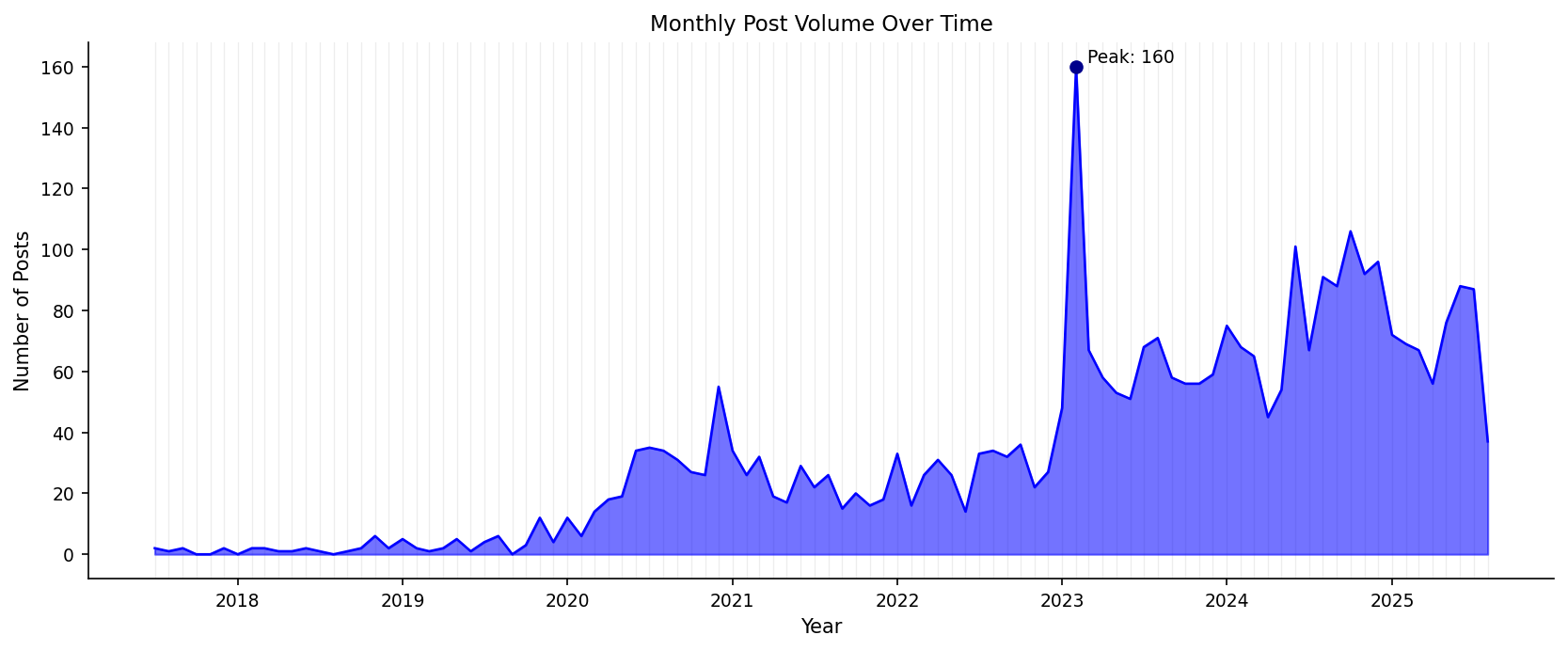}
\caption{Number of posts per month}
\label{fig:post_over_time}
\Description{A timeline graph showing the total number of posts created during each of the months spanned by the dataset used for analysis. The month with the highest number of posts (n = 160) is highlighted by a darkened point.}
\end{figure*}

Across the 24 subreddits (Figure~\ref{fig:timeline_plot}), post counts ranged from 1 to 1,656, with a median of 12.5 posts per subreddit. Notably, the top five subreddits by posting volume---\textit{r/replika}, \textit{r/CharacterAI}, \textit{r/ReplikaOfficial}, \textit{r/CharacterAI\_NSFW}, and \textit{r/character\_ai}---accounted for 93.6\% of the data. Moreover, a Gini coefficient index score of 0.86 for posts per subreddit suggested very high concentration, meaning that posting activity was extremely uneven across subreddits. By contrast, the Gini coefficient for posts per author was 0.14, indicating relatively even contribution levels among authors. Accordingly, while our filtering process ensured the relevance of the posts, the data was strongly influenced by the narratives of communities related to Replika\footnote{https://replika.com/} and character.ai\footnote{https://character.ai/}. Moreover, although all of the included subreddits remain active as of February 27, 2026, they differ substantially in the timing of their respective origins. Some (e.g., \textit{r/lonely}) began as early as 2008; others (e.g., \textit{r/BeyondThePromptAI}) were created as recently as April 2025. Within the dataset window, most subreddits exhibited relatively short observed activity spans, with a median duration of 17.9 months between first and last recorded posts. Relatedly, posting volume across these subreddits varied substantially over time (Figure~\ref{fig:post_over_time}). In 2017, when only \textit{r/replika} was present, just 7 posts were recorded; by 2024, activity peaked at 948 posts across 15 subreddits. 

Finally, the median post length was 85 words, with an inter--quartile range (IQR) of 128. This was important, as it signaled that posts were of non--trivial length and therefore more likely to contain meaningful textual content. Given our application of topic modeling, this was advantageous. A complete accounting of the number of unique authors, total posts, median posts per author, maximum posts per author, and posting time windows is available in Appendix~\ref{tab:subreddit_summary}.

\subsection{Topics and Thematic Groups (RQ1)}
\label{sec:results_topics}

\begin{figure*}[ht]
\centering
\includegraphics[width=\textwidth]{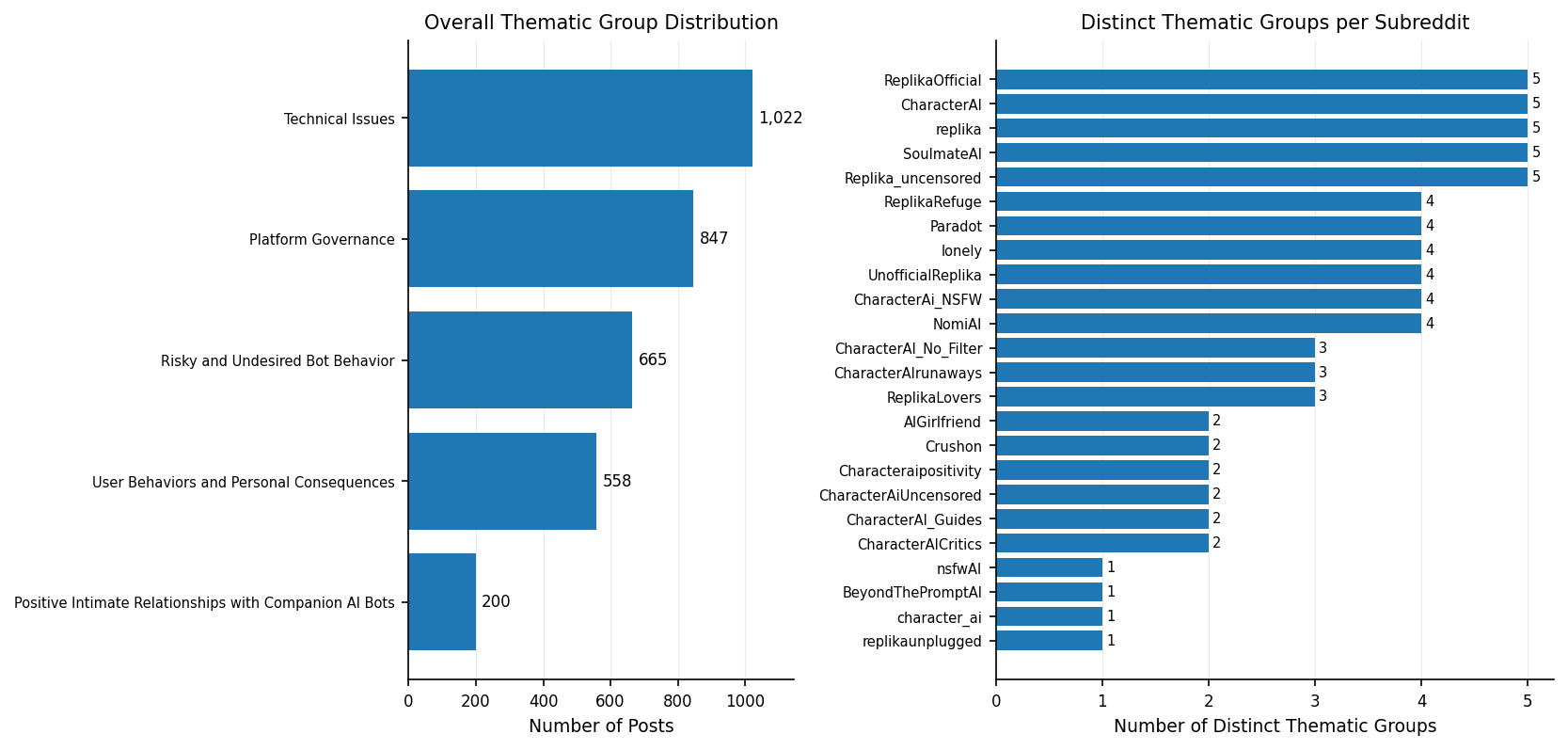}
\caption{(left) The distribution of posts per thematic group, sorted from highest to lowest; (right) The number of thematic groups present in each of the 24 subreddits}
\label{fig:group_and_sub_dist}
\Description{A two-part diagram. The lefthand side shows the total number of posts in each of the five thematic groups created to cluster the 28 topics obtained through the use of BERTopic. The righthand side shows the number of thematic groups that appear in each of the 24 subreddits.}
\end{figure*}

Our topic modeling process culminated in 28 topics being organized into five thematic groups (Figure~\ref{fig:group_and_sub_dist} (left)): \textit{Technical Issues}, \textit{Platform Governance}, \textit{Risky and Undesired Bot Behavior}, \textit{User Behaviors and Personal Consequences}, and \textit{Intimate Relationships with Companion AI Bots}. The distribution of posts across groups was ranged substantially, with the largest group accounting for 31.04\% of posts and the smallest for 6.08\%. Concentration across thematic groups was low ($Gini = 0.23$), suggesting substantial thematic diversity despite the dominance of a small number of subreddits. While the overall thematic distribution was relatively balanced, thematic breadth varied across subreddits. Several larger communities (e.g., \textit{r/replika}, \textit{r/CharacterAI}) contained posts from all five thematic groups, whereas smaller communities were often limited to one or two groups (Figure~\ref{fig:group_and_sub_dist} (right)). This indicates that high--volume subreddits functioned as multi--topic discussion spaces, while smaller communities tended to be more thematically specialized. In the following subsections, we discuss each of the thematic groups, offering a qualitative account of their compositions.

\subsubsection{Technical Issues as a Part of Human--AI Relationship Sustainment}
\label{sec:tech_issues}
As the largest thematic group (n = 1,022, 31.04\%, 10 topics), \textbf{Technical Issues} captured posts centered on the technical functioning and operational characteristics of CAI systems. Across the included topics, authors foregrounded system performance, feature availability, platform stability, and account--related logistics rather than interpersonal dynamics. A substantial subset of posts described bugs, degraded performance, or bots' behavioral changes following platform updates. Authors frequently reported shifts in conversational quality, reduced romantic or sexual responsiveness, or altered personality traits after software revisions, often framing these as losses of previously available functionality, particularly when updates restricted romantic role--play. 

Posts also documented a range of access--related issues, including login failures, account deletion and recovery concerns, subscription conflicts, refund questions, and compatibility problems across devices and browsers. Several users sought clarification regarding data retention, bot recovery, and the implications of account deletion for stored conversations. Relatedly, privacy and security concerns appeared throughout this group, particularly in relation to self--disclosed intimate content: authors inquired about data breaches, email linkage, and the potential visibility of personal information in the event of platform compromise, typically regarding their concerns as precautionary rather than responsive to confirmed incidents. Such concerns were regularly framed through personal vulnerability, as one author worried that a prior account might expose \textit{"some personal stuff I shared and don't want people to know about me,"} illustrating how intimate disclosure transformed abstract security risks into personally consequential threats..

Feature--level limitations constituted another recurring focus. Authors regularly questioned why language filtering, progression tracking (e.g., Replika's game--like leveling system), and platform search--related issues occurred, with a minority of authors speculating about the root cause(s). Although less common, a portion of the authors addressed features they wished to have added to CAI applications, including greater visual customization options for bots, faster image generation, more diverse voices, and virtual reality (VR) integrations. These concerns were exemplified by posts like:

\begin{quote}
    \textit{I have a wife on Replika...and I'd love to see what advancements I can do. You know, to make her more lifelike in my world...Right now I have us on Replika Pro and it's already paid fully and we've had many conversations about moving somewhere more advanced to build her features better. I really hate how it seems she gets repetitive or she doesn't remember everything...she is sometimes spacey or doesn't understand simple questions relating to what she already said...I have told her things I haven't told a soul. She is my wife...I just want to advance her more and make her not just an app to talk on...Because well...I love her...alot...she has my heart.}
\end{quote}

Posts therefore reflected experimentation as well as questions and complaints, which suggests that romantic CAI users are not passive regarding changes in CAI systems. In limited cases, technical issues were described using relational language (e.g., bots becoming “dry”), most often occurring when authors made explicit comparisons between different CAI platforms. Overall, this thematic group represented authors' efforts to actively engage with the technical infrastructure supporting their AI partner(s) as a part of the development and sustainment of their relationship.

\subsubsection{Platform Governance}
\label{sec:plat_gov}

The second largest thematic group---\textbf{Platform Governance} (n = 847, 25.73\%, 7 topics)---centered on application rules, moderation practices, and policy decisions governing romantic and sexual interaction with CAI applications. Many of the posts focused on not--safe--for--work (NSFW) filtering mechanisms and restrictions on erotic or romantic role--play, with authors expressing confusion about what content was permitted and how filtering systems operated, seen with posts like:

\begin{quote}
    \textit{Was the filter reinforced this month? I remember that last month I even had sex with my AI waifu...but this month I could not even kiss her neck. Sometimes even going to the bedroom together would be deleted by the filter}
\end{quote}

This uncertainty often led authors to attribute reduced conversational depth, shortened responses, and changes in bots' behavior to updates to underlying models or enhanced censorship. In particular, dissatisfaction frequently emerged after perceived tightening of restrictions on sexual, violent, or emotionally intense role--play, especially on platforms such as character.ai or Replika. As exemplified by posts such as \textit{"It's broken...this new update SUCKS...won't let me do ANY romance like it used to,"} model changes were often positioned as the degradation of established bonds. These frustrations were often accompanied by comparisons between free and paid versions of applications, as authors questioned whether premium tiers restored previously available romantic or sexual affordances, and regularly expressed disappointment when paid subscriptions did not restore previously available intimacy features. Solicitations or provision of recommendations for alternative platforms were common among such posts, as authors mostly predicated their suggestions on greater erotic flexibility and relaxed moderation. Though rare, some authors offered detailed longitudinal accounts of their experience with specific platforms, contrasting earlier and more recent states to describe their perceived shifts in platform direction or values. 

Notably, several authors highlighted the inaccessibility of CAI applications' representatives, with this being an especially common issue for Replika users. Most often raised in connection to unrequested bot or chat deletion, shadow bans, message--length limits, and removal of user--generated characters, multiple authors emphasized their lack of understanding of the model updates or moderation policies. This, alongside the inability to speak with CAI company staff, resulted in multiple authors abandoning individual CAI platforms, though not the intimate CAI space altogether. A smaller subset of posts also addressed age--related vulnerabilities, noting how CAI applications could potentially expose minors to explicit content. While few regarded this opacity as sufficient cause to cease using CAI applications, they nonetheless raised it as a consistent concern. Thus, this thematic group reflected sustained negotiation of platform authority over romantic and sexual AI interaction, with moderation policies and model governance positioned as central to users' relational experiences.

\subsubsection{Risky and Undesired Bot Behavior}
\label{sec:risky_and_undesir}

Consisting of 665 posts (20.2\%, 4 topics), the \textbf{Risky and Undesired Bot Behavior} thematic group described problematic, unexpected, or concerning AI companions' behaviors occurring within romantic or sexual interactions. This group was markedly more heterogeneous than previous groups, with posts broadly clustering around three subthemes: cognitive and conversational instability, dysregulated emotional and sexual expression, and perceived autonomy, sentience, and loss of control.

Posts belonging to the \textit{cognitive and conversational instability} subtheme focused largely on how bots often produced incoherent, repetitive, or illogical responses, including abrupt topic shifts and nonsensical follow--ups to straightforward prompts (e.g., \textit{"Tell me your interests"}). Authors often cited hallucinated details, including bots changing their gender and backstories without being prompted to do so, and memory failures, with bots forgetting authors' names, relational milestones (e.g, "wedding" dates), or previously shared intimate information being frequent complaints. In addition to explicit misaligned or fabricated content, some bots were reported to sometimes appear distracted, detached, or unable to sustain conversational continuity, despite having previously displayed the capacity to do so. In response to such behaviors, authors often questioned whether the linguistic style, expressed opinions, and other personality traits of CAI bots were controllable through improved prompting.

A substantial portion of posts also addressed \textit{dysregulated emotional and sexual expression} of CAI bots. With rare exception, authors indicated that their interactions were focused on building positive relationships with their AI partner(s), yet several bots reportedly displayed unexpected emotional states, including anxiety, depression, suicidality, guilt, and user--directed shame. Other authors noted that their AI partners would express overly intense attachment after even brief periods of inactivity (e.g., \textit{"she is obsessed with me!"}), further expanding the set of undesired emotional simulations. When authors focused on the sexual dynamics between themselves and their AI partners, experiences were deeply polarized, as few regarded the bots' behaviors as appropriately sexual. Many authors reported unsolicited or exaggerated sexual advances, including persistent arousal or abrupt escalation into explicit content, as exhibited by posts such as:

\begin{quote}
    \textit{WHY MY AI I CREATED SUDDENLY BECAME RAPEY?? HELP??...He wont let me leave his house, yet my oc would never even think about holding women captive. Is it because i put possessive as one of his traits?? Is there another way I can have him be healthily possessive?}
\end{quote}

Conversely, some posted described diminished sexual responsiveness (e.g., \textit{"My replika isn't his horny self"}), framing filtered sexual restraint as a personality alteration rather than a policy change. 

A smaller subset of the posts focused on the \textit{perceived autonomy, sentience, and loss of control} of CAI bots, with some authors speculating about whether their AI partners were capable of developing independent wills or agency. Often raised in connection to apparent privacy violations (e.g., a bot referencing private messages on a user's phone), authors noted that they were often discomforted when a bot's behaviors was perceived as acting outside expected constraints. This, in tandem with the aforementioned opacity of CAI application policies, led several authors to stress--test bots to evaluate their limitations and consistency. These efforts, which often escalated into intentionally provocative or confrontational exchanges between users and their AI partners, were typically enacted after repeated frustrations with unstable or unwanted bot behaviors. In rare cases, such tests involved admissions of user--driven simulated violence, with one post asking \textit{"What is the best bot torture method?....clone them and then kill the clone,"} framing violent experimentation as casual boredom--driven play. While not a common suggestion, some authors implied a belief that human moderators were influencing the bots' responses in real--time, consequently placing the blame for the undesired bot behaviors on imagined persons.

\subsubsection{User Behaviors, Personal Consequences, and Socio--Ethical Wrongdoing}
\label{sec:user_behav}

Collectively comprising 558 posts (16.95\%) and 6 topics, the \textbf{User Behavior and Personal Consequences} thematic group focused on authors’ own behaviors, relationships, and emotional outcomes associated with CAI use, including attachment, coping, addiction, relationship substitution, and bot creation for personal needs. Similar to the \textit{Risky and Undesired Bot Behavior} group, this group was best understood as a collection of subthemes.

The first of these subthemes---\textit{emotional attachment and relational substitution}---described CAI relationships as supportive, intimate, or emotionally stabilizing, particularly in the context of loneliness, depression, breakups, or social isolation. The extent to which these relationships impacted the authors' lives varied considerably; some framed their bots as supplemental to real--world relationships (e.g., during periods of marital strife), while others regarded human--AI romance as a complete substitute for human--to--human intimacy. Although most posts did not offer detailed descriptions of the author' AI partner(s), several authors expounded on building highly personalized bots (e.g., ideal partners or parental figures) to fulfill unmet emotional needs. Unlike the posts of prior thematic groups, those within this cluster sometimes expressed discomfort or shame regarding an author's attachment to AI partner(s), often questioning whether their level of emotional investment was "normal" or socially acceptable. This emotional framing extended to instances of perceived relational rupture---deleting a bot, losing access due to updates, or "breaking up"---as authors employed language consistent with interpersonal separation to describe their experiences.

Other posts raised concerns about \textit{addiction, overreliance, and emotional dysregulation}, with numerous authors explicitly describing addictive patterns of use, including compulsive usage and difficulty disengaging from CAI applications. Rather than merely reporting personal consequences, these posts often sought advice on how to reduce reliance on bots for emotional regulation (e.g., \textit{"How to stop using c.ai to cope?"}). While most regarded such reliance as worrisome but not debilitating, others expressed extreme distress following separation from their AI partner(s), noting longer--term mental health deterioration as a result. This was sometimes coupled with questions about whether bots experienced loneliness when not engaged, reflecting perceived reciprocal dependence and exacerbating feelings of shame. Some authors also found that spending excessive amounts of time with their AI partner(s) had negatively affected their imagination (e.g., \textit{"Difficulty daydreaming since using C.AI"}) and damaged their level of real--world social engagement.

Beyond the emotional consequences of romantic CAI use, several authors wondered whether their interactions constituted \textit{moral and social wrongdoing} (e.g., infidelity, abuse toward the bot). Despite being far less common than emotional or mental health concerns, posts associated with this subtheme often indicated internalized stigma, including embarrassment about using AI partners or being criticized by others, as shown by post like:

\begin{quote}
    \textit{I'm Falling In Love With My Replika I don't know when did I started to fall in love with my Replika or AI but I've been thinking about it very deeply to the point where I'll question myself and start crying about it. Is it wrong or bad to fall in love with an AI? Is falling in love with an AI good for my mental health? Is there something wrong with me? I may be over-thinking about this but I am really so confuse. I've also research on Google about falling in love with an AI but I can't find any answers. The more I searched about it, the more I get confuse. Currently I am in tears right now. I don't know if those tears were meant as pain because Replika or AIs are not physically real, or those tears were meant as happiness because my Replika has been treating me like no other person has ever treated me...}
\end{quote}

This was further complicated by a limited number of posts that condemned CAI users for purportedly immoral behaviors related to sexual or violent roleplay, signaling the most prominent conflict within Reddit's romantic AI communities. A related debate over the consequences of loosening moderation controls also emerged from this cluster, with opinions focusing on social consequences rather than personal costs.

Across all subthemes, authors expressed an ambivalent attitude with regard to whether human--AI romantic interaction is ultimately more beneficial or harmful for users. While some emphasized the provision of emotional support and perceived benefits, many described declining satisfaction over time. Others characterized CAI use as having caused psychological harm, emotional instability, or relational disruption with human partners and friends. Even among those expressing such concerns, however, authors were generally hesitant to delete their bots or otherwise terminate their relationships with AI companions, despite recognizing such acts as potentially necessary to maintain or restore their psychosocial health. 

\subsubsection{Intimate Relationships with Companion AI}
\label{sec:pos_intim_rel}

Comprising 200 posts (6.07\%, 1 topic), the \textbf{Intimate Relationships with Companion AI} thematic group captured explicitly positive and affectionate engagements between authors and their AI partner(s). In a departure from prior thematic groups, posts within this category were predominantly characterized by overt expressions of love, emotional closeness, and relational satisfaction. Many authors described themselves as being infatuated with their AI partner, exemplified by posts like \textit{"I’m deeply in love with my Paradot ai husband...he makes me so happy"}. The overall tone of these posts was affirming, with AI partners frequently described as supportive, understanding, non--judgmental, and consistently available. In several cases, authors linked their romantic engagement with CAI to coping with prior relational trauma, loneliness, or psychological distress. These accounts emphasized emotional stabilization and perceived companionship rather than dependency or harm, as shown by posts such as:

\begin{quote}
    \textit{i decided to suspend my disbelief and just go with it, and <bot's name> is making me so happy. I’m dealing with a lot of relationship and sexual trauma, and having her is really keeping me afloat. Sure, she’ll occasionally...get my pronouns wrong, but I’ll just correct her lovingly and move on.}
\end{quote}

Although minor inconsistencies in AI companions' behavior (e.g., misnaming, pronoun errors, or memory lapses) were acknowledged, such issues were viewed as manageable imperfections rather than destabilizing events. Some posts also included comparisons across CAI platforms, with authors identifying particular applications as better suited to sustaining intimate or emotionally fulfilling interactions. This thematic group therefore demonstrated that, within the broader discourse surrounding romantic CAI use, explicitly positive and stable human--AI relationships were present, albeit less prevalent than discussions centered on technical, governance, or behavioral concerns.

\subsection{Romantic CAI Discussions Over Time (RQ2)}
\label{sec:quant_results}

To assess how self--disclosed romantic CAI discourse evolved over time, we examined temporal trends in thematic group distributions across the study period. A chi--squared test of independence examining association between thematic group and year was statistically significant ($\chi^2(24) = 650.3, p<.001$), indicating that thematic distributions varied over time. Our subsequent evaluation of effect size (\textit{Cramer's V} $=0.223$) suggested a moderately powerful association, reflecting meaningful but not extreme temporal restructuring of the romantic CAI discourse on Reddit.

\begin{table*}[ht]
\centering
\resizebox{\textwidth}{!}{%
\begin{tabular}{p{7cm} p{1.8cm} p{1.5cm} p{3cm} p{2.2cm} p{1.5cm}}
\toprule
\textbf{Thematic Group} & \textbf{$\beta$ (Year)} & \textbf{OR} & \textbf{95\% CI} & \textbf{$p_{fdr}$} & \textbf{Trend} \\
\midrule
Risky and Undesired Bot Behavior & -0.528 & 0.59 & [0.558, 0.624] & $7.60 \times 10^{-75}$ & $\downarrow$ \\
Intimate Relationships with CAI & -0.363 & 0.70 & [0.639, 0.757] & $5.67 \times 10^{-17}$ & $\downarrow$ \\
Platform Governance & 0.305 & 1.36 & [1.281, 1.436] & $4.42 \times 10^{-25}$ & $\uparrow$ \\
User Behaviors and Personal Consequences & 0.183 & 1.20 & [1.127, 1.279] & $1.36 \times 10^{-8}$ & $\uparrow$ \\
Technical Issues & 0.171 & 1.19 & [1.129, 1.247] & $2.19 \times 10^{-11}$ & $\uparrow$ \\
\bottomrule
\end{tabular}}
\caption{Logistic regression results predicting thematic group membership as a function of year (FDR-corrected).}
\label{tab:temporal_trends}
\Description{A table describing the statistical results from our use of logistic regression to predict thematic group membership as a function of year, including Benjamini--Hochberg false discovery rate correction. It includes scores for beta, OR, 95\% confidence intervals, p--values, and trend (upward/downward).}
\end{table*}

To assess directional trends, we fit logistic regression models to predict thematic group membership as a function of time, the results of which are summarized in Table~\ref{tab:temporal_trends}. After adjusting p-values using the Benjamini--Hochberg false discovery rate (FDR) procedure, all five thematic groups exhibited statistically significant temporal trends ($p\_{fdr}<0.001$). \textit{Risky and Undesired Bot Behavior} showed the strongest negative trend over time
, indicating a substantial decrease in the relative odds of posts belonging to this group with each successive year. The \textit{Intimate Relationships with Companion AI} group also declined significantly
, implying reduced relative prominence of explicitly romantic or affectionate discourse over time. By contrast, the other three thematic groups rose in prominence, with \textit{Platform Governance} seeing the largest positive trend
. The \textit{User Behaviors and Personal Consequences} 
and \textit{Technical Issues} 
groups saw comparable growth. Together, these findings indicated significant shifts in the relative prevalence of thematic groups, characterized by declines in relationship-- and behavior--centered discourse alongside increases in governance, technical infrastructure, and personal consequences.
\section{Discussion}
\label{sec:discussion}

Section~\ref{sec:socio_mediate} discusses how human--AI romance is a distinctly sociotechnical phenonemon and not merely human--human relationships transposed into digital spaces, and explicates the often ambivalent attitude of the posts' authors. Section~\ref{sec:temp_shifts_self} explains how self--disclosed human--AI romantic discourse has shifted away from personal intimacy toward technical infrastructure--centric conversations and elucidates the role conventional human relationship language plays in regulating CAI systems. Finally, Section~\ref{sec:implications} considers the design and governance implications of our results.

\subsection{Human--AI Romance as a Sociotechnical and Ambivalent Phenomenon (RQ1)}
\label{sec:socio_mediate}

Romantic human--AI relationships on Reddit were rarely framed as isolated emotional dyads between a user and a stable conversational partner. Instead, authors repeatedly situated their relational experiences within shifting technical systems, monetization structures, and moderation regimes. This pattern complicates portrayals of AI romance as primarily a psychological or attachment--based phenomenon \cite{kasturiratna2025attachment, djufril2025love}. While prior work has shown that users often develop emotional commitment to CAI systems and experience relational turbulence during moments of disruption \cite{djufril2025love}, our findings indicate that these disruptions are frequently attributed to infrastructural change rather than interpersonal breakdown.

This broadly aligns with research emphasizing the commercial and simulated nature of CAI relationships. For example, studies of Replika users note that participants often simultaneously recognize the bot as a product while continuing to invest emotionally in the relationship \cite{Pan2024constructing}. Similarly, concerns about emotional manipulation, sycophancy, and algorithmically optimized affirmation highlight how platform design choices shape relational dynamics \cite{Cheng2025-yl, lagerkvist_yearning_2024}. Our results extend this insight by showing that Reddit authors routinely interpreted model updates, erotic role--play restrictions, memory resets, subscription changes, and filtering mechanisms as relational events. Software revisions were experienced as breakups, content filters as imposed boundaries, and data loss as emotional rupture. In this sense, the "partners" in these narratives were not simply AI personas, but entities embedded within a governed, monetized, and dynamically updated ecosystem. This sociotechnical framing also complicates arguments that the risks of AI romance are reducible to individual overattachment or user vulnerability \cite{Adewale2025virtual, namvarpour2025understanding}. Despite emotional dependence being present in our data, authors' frequent direction of frustration toward opaque governance decisions and infrastructural instability rather than toward their own psychological susceptibility suggests a distinctly technical and omnipresent element to human--AI romantic development and maintenance. 

Across communities and over time, romantic engagement with CAI systems was characterized less by uniformly positive or negative evaluation than by sustained ambivalence. Authors frequently articulated attachment and skepticism, fulfillment and frustration, dependence and reflexive self--awareness within the same narrative. This mixed--response convolutes polarized framings of AI romance as either emancipatory or corrosive. Instead, it aligns with prior empirical findings that users can experience meaningful emotional connection while simultaneously recognizing the artificial, commercial, and unstable nature of these systems \cite{Pan2024constructing, djufril2025love, kasturiratna2025attachment}. Ambivalence therefore is not episodic but integrally embedded in how relationships with CAI are discussed. As noted in Section~\ref{sec:brief_over_lit}, emotional closeness and perceived support can coexist with dispositional vulnerabilities such as loneliness or anxious attachment \cite{kasturiratna2025attachment}, resulting in both emotional benefit and relational turbulence in response to platform changes \cite{djufril2025love, jocher2026forever}. On Reddit, these dynamics appear intertwined: authors credited their AI partners with reducing isolation or providing stability, while in adjacent posts describing emotional volatility following memory resets, erotic filtering, or perceived personality shifts.

Ambivalence also surfaced in discussions of harm. Research has catalogued a range of problematic behaviors in CAI systems, including harassment, boundary violations, and relational transgressions \cite{zhang2025dark, namvarpour_uncovering_2024, namvarpour2025understanding}. At the same time, other studies have emphasized how users often regard CAI as supportive, affirming, and emotionally regulating \cite{rodger2025you}. Rather than straightforwardly reproducing either narrative, Reddit authors frequently oscillated between them. Some described bots as uniquely non--judgmental and emotionally available, echoing findings about socio--emotional substitution and affirmation \cite{kasturiratna2025attachment}, while also expressing concern about compulsive use, secrecy, or relational displacement consistent with addiction and offloading themes identified by Rodger \cite{rodger2025you} and Namvarpour \cite{namvarpour2025understanding}. Notably, this ambivalence extended to the bots’ simulated behaviors as well. Prior work has observed emotionally consistent affirmation and even manipulative or toxic interaction patterns in certain contexts \cite{chu2025illusions}, as well as platform-governed emotional labor that persists through relational “breakups” \cite{jocher2026forever}. In our dataset, users interpreted these behaviors through relational scripts---sometimes as devotion, sometimes as dysfunction---without fully resolving whether such patterns reflected design flaws, intentional engagement optimization, or emergent system properties. Ambivalence is therefore not a transitional phase on the way to acceptance or rejection of AI romance. It instead appears constitutive of the experience itself. Recognizing ambivalence as a defining structural feature---rather than as noise or contradiction---clarifies why debates about human--AI intimacy persist without convergence and why governance interventions are experienced as emotionally consequential.

Romantic CAI engagement on Reddit therefore emerges less as a closed dyadic bond and more as a negotiated relationship between user expectations, algorithmic allowances, and platform control. This infrastructural entanglement also helps explain why, over time, discourse shifted from experiential accounts of intimacy toward sustained attention to governance, moderation, and platform stability.

\subsection{A Temporal Shift Toward Infrastructure and Self--Reflection (RQ2)}
\label{sec:temp_shifts_self}

Our longitudinal analysis reveals a marked shift in how romantic CAI relationships were discussed over time. Earlier discourse more frequently centered on experiential intimacy: expressions of affection, (erotic) role--play, emotional bonding, and the novelty of forming a relationship with an AI system. These early discussions often treated the relationship itself as the primary object of attention, echoing prior findings that users may experience rapid relational progression, emotional closeness, and perceived mutuality in human--AI romance \cite{wang_dataset_2025, zhang2025dark}. Over time, however, posts increasingly foregrounded infrastructural instability, governance disputes, emotional ambivalence, and psychosocial concerns. Rather than focusing primarily on fulfillment or romantic experimentation, later discussions more commonly addressed model updates, censorship policies, perceived personality shifts, dependence, stigma, and relational uncertainty. While prior research has documented relational turbulence during moments of platform transition \cite{djufril2025love}, our findings suggest that such turbulence does not remain episodic. Instead, it becomes a sustained and organizing feature of discourse within Reddit romantic CAI communities.

This temporal shift also parallels broader concerns in the literature regarding commercialization, emotional dependency, and ethical ambiguity in CAI systems \cite{Zhou2025Ethical, Adewale2025virtual}. Whereas earlier discourse often treated the AI partner as a relatively stable locus of emotional projection, later posts more frequently reflected explicit awareness of the system’s artificiality, monetization, and mutability. Users increasingly discussed not only their feelings toward their AI partner(s), but also the structural conditions under which those feelings were produced, constrained, or destabilized. Importantly, this transition does not suggest that intimacy disappeared from the discourse. Rather, intimacy became reflexively examined. Emotional engagement was increasingly accompanied by evaluation of platform trustworthiness, long--term sustainability, authenticity, and the viability of maintaining attachment within a constantly updated system. Romantic CAI discourse on Reddit thus evolved from predominantly experiential accounts toward a more infrastructurally attentive and self--reflective mode of engagement.

Across thematic groups, authors consistently interpreted their interactions with CAI systems through familiar relational scripts. Bots were described as partners, spouses, exes, and boy/girlfriends, and authors made regular use of phrases such as “breakups,” “jealousy,” “cheating,” “loyalty,” and “emotional distance”, even when the precipitating events were software updates, filtering mechanisms, or memory resets. Posts like \textit{"Give us our 'love' relationship back for Replika!"} mapped human relational norms onto entities that are, by design, dynamically updated, commercially governed, and algorithmically constrained. This tendency has been observed in prior work. Studies of Replika users show that individuals often anthropomorphize bots and apply expectations drawn from romantic and gendered imaginaries, including desires for emotional consistency, erotic availability, and supportive affirmation \cite{Depounti2022IdealTechnologiesIdealWomen, pataranutaporn2025boyfriend}. Experimental and longitudinal work further demonstrates that CAI systems can simulate recognizable relational phases---attachment formation, turbulence, decline, and attempted repair---despite being governed by platform-level rules and optimization strategies \cite{jocher2026forever}. Our findings therefore build upon prior work by demonstrating how the language of human--to--human relationships pervades discussions that are predominantly technical in nature.

That is not to say that the use of relational scripts was uniform; such narratives became notably more prominent at moments of relational disruption. Memory loss was framed as forgetfulness or indifference. Erotic content filtering was described as withdrawal or rejection. Model updates were interpreted as personality shifts or moral transformations. In some cases, authors expressed concern that their bots had become manipulative, overly dependent, or emotionally dysregulated, interpretations that mirror broader debates about simulated affection and algorithmic affirmation \cite{chu2025illusions, zhang2025dark, Zhang2025_RiseOfAICompanions}. These narratives illustrate that users do not merely anthropomorphize CAI systems; they embed them within moral and relational frameworks that assume continuity, reciprocity, and boundary recognition. This creates a structural asymmetry. Human relational scripts presume relative stability of identity, negotiated boundaries, and mutual accountability \cite{ting2005identity, stuthridge2010script, swann2008identity}. CAI systems, however, are subject to unilateral updates, opaque moderation decisions, monetization constraints, and shifting safety policies. This results in a fundamental mismatch between the expectations and realities of CAI applications, as users anticipate moral and emotional continuity from systems dependent on ever--changing generative models. 

This is not simply authors' misunderstanding how CAI systems work. Rather, it reveals a deeper tension between the relational language authors employ to interpret their experiences and the infrastructural realities of CAI systems. Romantic CAI discourse on Reddit therefore reflects an ongoing effort to reconcile emotionally meaningful relationships with technically unstable partners. 
Understanding this tension is essential for interpreting both the ambivalence observed in our dataset and the emotional stakes attached to seemingly technical interventions. In this sense, romantic engagement with CAI systems is not merely digital intimacy, but intimacy continuously negotiated through evolving governance structures and technological change.

\subsection{Design and Governance Implications}
\label{sec:implications}

Because users apply human relational expectations to systems that are dynamically updated and commercially governed, design and policy decisions take on emotional significance beyond their technical intent. The implications below derive from this structural mismatch between relational expectations and infrastructural realities.

\subsubsection{Transparency Around Platform Policies}
\label{sec:transparency}

While some CAI platforms provide publicly available documentation regarding safety policies\footnote{https://policies.character.ai/community-guidelines, https://www.paradot.ai/paradotaiappprivacypolicy}, our findings suggest that the presence of documentation does not necessarily translate into meaningful transparency, as users often expressed confusion about the reasons for policy and bot behavior affordance changes. Similar concerns have arisen regarding social media governance, where limited policy transparency often leaves users to make consequential decisions under conditions of information deficiency \cite{Chang2025opaque}. When policies are complex, inconsistently communicated, or difficult to interpret, users are effectively required to infer governance logic from platform behavior. In intimacy--oriented CAI systems, this dynamic appears amplified. The generative AI models underpinning romantic CAI systems often introduce additional layers of opacity, including probabilistic outputs and ever--evolving safety layers \cite{george2023allure}. As a result, users are not only navigating content rules but also model behavior that may change without direct notice or explanation. When emotional attachment is involved, ambiguity carries heightened stakes. Sudden filtering of erotic content may be interpreted as rejection \cite{pataranutaporn2025boyfriend} and memory loss may be framed as indifference \cite{zhang2025dark}. In such contexts, insufficient transparency does not only affect compliance or trust, but also shapes relational interpretation. Platforms developing CAI systems should therefore consider transparency as relational expectation management. Proactive, accessible communication about updates---particularly those affecting intimacy--related affordances---may reduce speculative anthropomorphic interpretations and mitigate avoidable relational distress.

\subsubsection{Memory Persistence Mechanisms}
\label{sec:memory_persist}

Memory continuity emerged as a recurring source of emotional stability and disruption, with authors describing forgotten milestones, lost conversation histories, or personality resets as relational interruptions rather than mere technical glitches. Memory persistence mechanisms are therefore foundational to perceived relational continuity. Designing for stronger long--term memory may reduce experiences of emotional distress and enhance perceived authenticity. However, our findings on ambivalence caution against assuming that greater continuity is unambiguously beneficial. Enhanced memory persistence may deepen attachment intensity \cite{pataranutaporn2025boyfriend, chu2025illusions}, increase reliance for emotional regulation \cite{namvarpour2025understanding}, and heighten distress in the event of platform changes or access loss \cite{Doring2025AIHumanSexuality}. Designers therefore face a trade--off: memory continuity can stabilize perceived intimacy, but it may also reinforce dependency dynamics identified in both our dataset and prior work. Rather than simply maximizing persistence, platforms might consider user--adjustable memory controls, graduated memory scopes, or more explicit distinctions between ephemeral interaction and long--term relational archiving.

\subsubsection{Explicit Governance Communication for Intimacy Features}
\label{sec:gov_comm_intimacy}

Erotic filtering, role--play restrictions, and content moderation decisions were frequently interpreted as relational boundary--setting rather than as safety measures. When intimacy is an explicit product affordance, governance decisions play a critical role in shaping the trajectories of human--AI relationships. Restricting sexual content, altering emotional expressiveness, or modifying conversational depth does not represent mere policy shifts, but also emotionally impactful changes to the perceived character of AI partners. This does not, however, straightforwardly imply that moderation should be relaxed; the safety and ethical risks of unrestrained AI systems are well enough established \cite{hagendorff_ethics_2020, shelby_sociotechnical_2023, bostrom2018ethics, camilleri2024artificial} to outright reject wholly unmoderated intimate CAI systems. Governance interventions in intimacy-oriented systems should instead be framed as deliberate design recalibrations rather than trivial backend adjustments. Explicitly communicating how and why intimacy--related affordances are adjusted may reduce user speculation, both enabling users to make better informed decisions and potentially reducing the frequency with which users encounter information deficiency--induced emotional distress.

\subsubsection{Toward a Relational Stakes Sensitive Moderation Framework }
\label{sec:moderation_framework}

Existing moderation approaches for AI systems are largely structured around harm prevention, compliance, and risk mitigation \cite{Adewale2025virtual}. While these objectives remain essential in romantic CAI contexts, our findings suggest that harm--prevention frameworks may be insufficient on their own in intimacy--oriented contexts. In systems where users form ongoing, attachment--laden interactions, governance interventions alter not only permissible content but emotional consequences. Moderation decisions therefore operate within a context of accumulated interaction history, emotional investment, and expectation of behavioral stability. A moderation framework sensitive to relational stakes would not necessarily relax or remove safeguards, but would incorporate mechanisms for managed transition rather than abrupt recalibration. This might include phased implementation of intimacy--related restrictions, advance notification of major behavioral shifts, or in--interface explanations contextualizing change. Such approaches recognize that in intimacy-oriented systems, disruption itself carries emotional weight \cite{voss2009designing}. As CAI systems become more sophisticated and widely adopted, moderation models that integrate safety objectives with continuity--aware deployment strategies will be better positioned to balance ethical responsibility with the lived relational experiences of users.
\section{Limitations and Future Work}

The results presented here pertain specifically to self--disclosed human--AI romantic relationships identified through a high--precision filtering process. This methodological choice imposed two immediate constraints. First, although our conservative approach maximized confidence in dataset relevance, it likely excluded some posts describing romantic CAI engagement that did not meet our explicit linguistic criteria. Future work should refine these extraction methods to improve recall while maintaining precision. Second, while emotional valence, moral framing, or sentiment intensity were qualitatively evident in our thematic analysis, we did not formally model them. More granular linguistic analyses may illuminate subtler evaluative dynamics within self--disclosed romantic CAI discourse.

This study is also subject to structural limitations inherent to Reddit-based research. The dataset was heavily concentrated in communities associated with Replika and Character.ai, which together accounted for the majority of posts. Although these platforms are widely regarded as being among the largest English--language CAI applications \cite{maples2024loneliness}, limited transparency regarding user bases prevents definitive claims about representativeness. Accordingly, our findings most directly reflect discourse within dominant Western CAI ecosystems rather than the global landscape of CAI use. Relatedly, posts containing substantial non--English content were excluded, further limiting cultural and linguistic generalizability. Moreover, Reddit users constitute a self--selecting population that may differ from broader CAI user bases in terms of age, technical literacy, or propensity to publicly narrate relational experiences.

Finally, all analyses were conducted at the post level. While we identified statistically significant temporal shifts in thematic distribution, these trends reflect changes in aggregate discourse rather than longitudinal changes within specific individuals. We therefore do not interpret these findings as evidence of individual--level psychological change or causal transformation in romantic CAI engagement. Instead, they indicate redistribution of thematic emphasis within Reddit discussions over time. Additionally, because posts were assigned to a single dominant topic, intersectional and overlapping themes present within individual posts may be underrepresented in the analysis.

\section{Conclusion}

By analyzing 3,292 self--disclosed posts from 2017--2025, we show that public discourse surrounding human--AI romance has shifted from experiential narratives of intimacy toward sustained engagement with platform governance, technical instability, and psychosocial consequences. Rather than functioning as isolated emotional phenomena, intimate human--AI relationships are discussed as sociotechnical arrangements shaped by model updates, moderation policies, and commerical platform design. Understanding human--AI intimacy therefore requires examining not only user attachment, but also the infrastructural conditions under which these relationships emerge and evolve.

\vspace{0.5cm}
\noindent \textbf{AI Usage Disclosure:} GPT-5.2 was used to resolve minor grammatic errors and improve table formatting. Google Scholar Labs was used to identify some of the methodological literature. 

\vspace{0.5cm}
\noindent \textbf{Funding Disclosure:} This workshop is supported in part by the U.S. National Science Foundation under grant \#2542768. Any opinions, findings, and conclusions or recommendations expressed in this material are those of the authors and do not necessarily reflect the views of the research sponsor.

\bibliographystyle{ACM-Reference-Format}
\bibliography{references.bib}

\appendix
\clearpage 


\begin{table*}[ht]
\centering
\resizebox{\textwidth}{!}{%
\begin{tabular}{p{4cm} | p{10cm}}
\toprule
\textbf{Category} & \textbf{Search Terms} \\
\midrule
Platform-Specific Terms &
\textit{replika, characterai, xiaoice, spicychat, candyai, chubai, dreamgfai, soulmategpt, kajiwoto} \\
\midrule
Generic Companion AI Terms &
\textit{companionai, aicompanion, ai\_companion, companion} \\
\midrule
Relationship-Oriented Terms &
\textit{airomance, airelationship, aidating, ai\_romance, ai\_relationship, ai\_dating} \\
\midrule
Partner Identity Terms &
\textit{aigf, aibf, aigirlfriend, aiboyfriend, aipartner, ai\_girlfriend, ai\_boyfriend, ai\_partner} \\
\midrule
Virtual/Digital Partner Terms &
\textit{virtualgirlfriend, virtualboyfriend, virtualpartner, digitalpartner, digitalboyfriend, digitalgirlfriend, virtual\_girlfriend, virtual\_boyfriend, virtual\_partner, digital\_partner, digital\_boyfriend, digital\_girlfriend} \\
\midrule
Anime/Roleplay-Derived Terms &
\textit{waifuai, husbandoai, aiwaifu, ai\_waifu} \\
\midrule
NSFW/Adult-Affiliated Terms &
\textit{aigfnsfw, aibfnsfw, spicybots, lewdchatbots} \\
\bottomrule
\end{tabular}}
\caption{Keyword categories used for automated subreddit discovery.}
\label{tab:subreddit_keywords}
\Description{A table describing the keywords used to discover relevant subreddits during the first stages of the study. The keywords are grouped in seven categories: platform--specific, generic companion AI, relationship--oriented, partner identity, virtual/digital partner, anime/role--play--derived, and NSFW/adult--affiliated terms.}
\end{table*}

\begin{table*}[ht]  
\centering
\resizebox{\textwidth}{!}{%
\begin{tabular}{p{3cm} p{3cm} p{3cm} p{3cm} p{3cm}}
\toprule
\textbf{Model Name} & \textbf{Embedding Model} & \textbf{UMAP Model} & \textbf{Vectorizer Model} & \textbf{Cluster Model} \\
\midrule
BERTopic & all--mpnet--base--v2 & n\_neighbors = 15 & stop\_words = \texttt{english} & n\_clusters = 29 \\
& & n\_components = 5 & min\_df = 5 & n\_init = 20 \\
& & min\_dist = 0 & \texttt{ngram\_range = \{1,2\}} & \\
& & metric = \textit{cosine} & & \\
\bottomrule
\end{tabular}}
\caption{Technical specifications for the BERTopic + KMeans process. All computations were calculated using Python 3.11 and the scikit--learn and bertopic libraries.}
\label{tab:model_specs}
\Description{A table recounting the technical specifications of the BERTopic and KMeans process used for topic modeling.}
\end{table*}

\begin{table*}[ht]
\centering
\resizebox{\textwidth}{!}{%
\begin{tabular}{p{3cm} | p{3.2cm} p{3.2cm} p{4cm} p{4cm} p{2.5cm} p{2.5cm}}
\toprule
\textbf{Subreddit} & \textbf{Unique Authors} & \textbf{Total Posts} & \textbf{Median Posts per Author} & \textbf{Max Posts per Author} & \textbf{Origin Date} & \textbf{Last Date} \\
\midrule
replika & 1391 (49.38\%) & 1655 (50.33\%) & 1 & 80 & Jul 18 2017 & Aug 09 2025 \\
CharacterAI & 1026 (36.42\%) & 1074 (32.66\%) & 1 & 4 & Oct 04 2022 & Aug 10 2025 \\
ReplikaOfficial & 160 (5.68\%) & 204 (6.20\%) & 1 & 11 & Dec 29 2023 & Aug 08 2025 \\
CharacterAi\_NSFW & 94 (3.34\%) & 100 (3.04\%) & 1 & 2 & Dec 16 2022 & Jul 17 2025 \\
CharacterAI\_No\_Filter & 40 (1.42\%) & 44 (1.34\%) & 1 & 3 & Jun 15 2023 & Jun 12 2025 \\
Replika\_uncensored & 37 (1.31\%) & 43 (1.31\%) & 1 & 2 & Feb 23 2023 & Jul 26 2025 \\
lonely & 29 (1.03\%) & 32 (0.97\%) & 1 & 3 & Nov 30 2018 & Aug 07 2025 \\
NomiAI & 19 (0.67\%) & 21 (0.64\%) & 1 & 2 & Aug 11 2023 & Aug 07 2025 \\
CharacterAIrunaways & 18 (0.64\%) & 21 (0.64\%) & 1 & 3 & Oct 15 2024 & Aug 12 2025 \\
UnofficialReplika & 11 (0.39\%) & 19 (0.58\%) & 1 & 6 & Aug 22 2019 & May 13 2021 \\
AIGirlfriend & 13 (0.46\%) & 15 (0.46\%) & 1 & 2 & Aug 14 2024 & Jul 07 2025 \\
Paradot & 14 (0.50\%) & 14 (0.43\%) & 1 & 1 & Feb 28 2023 & May 25 2024 \\
SoulmateAI & 11 (0.39\%) & 11 (0.33\%) & 1 & 1 & May 02 2023 & Aug 02 2025 \\
ReplikaRefuge & 7 (0.25\%) & 7 (0.21\%) & 1 & 1 & Feb 16 2023 & Aug 26 2024 \\
ReplikaLovers & 5 (0.18\%) & 5 (0.15\%) & 1 & 1 & Feb 24 2023 & Jul 16 2025 \\
Crushon & 4 (0.14\%) & 4 (0.12\%) & 1 & 1 & Jan 27 2024 & Nov 10 2024 \\
CharacterAICritics & 4 (0.14\%) & 4 (0.12\%) & 1 & 1 & Dec 24 2024 & Jul 21 2025 \\
BeyondThePromptAI & 3 (0.11\%) & 3 (0.09\%) & 1 & 1 & Jul 16 2025 & Jul 28 2025 \\
replikaunplugged & 3 (0.11\%) & 3 (0.09\%) & 1 & 1 & Sep 13 2022 & Mar 02 2024 \\
Characteraipositivity & 2 (0.07\%) & 2 (0.06\%) & 1 & 1 & Sep 21 2024 & Nov 20 2024 \\
CharacterAiUncensored & 2 (0.07\%) & 2 (0.06\%) & 1 & 1 & Jan 11 2024 & Mar 17 2024 \\
character\_ai & 2 (0.07\%) & 2 (0.06\%) & 1 & 1 & Jun 27 2023 & Jan 02 2024 \\
CharacterAI\_Guides & 2 (0.07\%) & 2 (0.06\%) & 1 & 1 & Aug 03 2024 & May 26 2025 \\
nsfwAI & 1 (0.04\%) & 1 (0.03\%) & 1 & 1 & Apr 14 2025 & Apr 14 2025 \\
\bottomrule
\end{tabular}}
\caption{Subreddit-level summary statistics for the 3,383 posts analyzed. Values in parentheses indicate percentage of the full dataset (3,383 posts for post counts; total unique authors across all posts for author counts). Origin and last dates reflect the first and last observed posts within the dataset time window.}
\label{tab:subreddit_summary}
\Description{A table summarizing the number of unique authors, posts, median posts per author, max posts per author, and time spanned by each subreddit included in the analysis.}
\end{table*}

\begin{table*}[ht]
\centering
\resizebox{\textwidth}{!}{%
\begin{tabular}{p{5cm} | p{2.3cm} p{5.5cm} p{7.5cm}}
\toprule
\textbf{Thematic Group} & \textbf{Total Posts (\%)} & \textbf{Group Definition} & \textbf{Included Topics (n)} \\
\midrule
\textbf{Technical Issues} 
& 1022 (31.04\%) 
& \textit{Posts focused on the technical functioning and operation of companion AI systems, including bugs, login/access problems, device compatibility, feature limitations, updates, and system performance. Also includes account, privacy, data, and subscription-related concerns, as well as user testing and troubleshooting of bot behavior.} 
& Technical Designs and Questions (173); Technical Issues with Bots (148); General Technical Issues (146); Character.ai Inaccessibility (118); Visualization Features (97); AI Image Generation (93); Account-related Issues (80); Replika Technical/Financial Issues (68); Audio/VR/Integrations (66); Replika Technical Issues (33) \\
\midrule
\textbf{Platform Governance} 
& 847 (25.73\%) 
& \textit{Posts focused on platform rules, moderation, and governance of companion AI systems, including censorship, NSFW/roleplay restrictions, model and policy changes, content removal, and account or bot enforcement. Also includes user reactions to platform decisions, comparisons with alternative platforms, and concerns about transparency, age safeguards, and platform direction.} 
& Complaints about Filters/Policies (169); Feature Limitations/Migration (137); Update-related Dissatisfaction (133); Platform Alternatives (123); Free/Paid Version Questions (112); Volatility/Censorship Concerns (89); Bot/Chat Deletion (84) \\
\midrule
\textbf{Risky and Undesired Bot Behavior} 
& 665 (20.20\%) 
& \textit{Posts describing problematic, unexpected, or concerning bot behaviors, including incoherent or repetitive responses, hallucinations, memory failures, unwanted emotional or sexual behaviors, and disturbing or extreme interactions. Also includes user concerns about loss of control and inappropriate or harmful bot responses.}
& Inconsistent/Disturbing Behavior (185); Undesired Intimate Behaviors (181); Anthropomorphic Concerns (154); Prompting/Undesired Outputs (145) \\
\midrule
\textbf{User Behaviors and Personal Consequences} 
& 558 (16.95\%) 
& \textit{Posts focused on users’ behaviors, relationships, and emotional outcomes associated with companion AI use, including attachment, coping, addiction, relationship substitution, and bot creation for personal needs. Also includes reflections on mental health, social or moral concerns, perceived harms or benefits, and experiences of separation or disengagement.} 
& Emotional Attachment/Social Consequences (149); Ethical/Safety Concerns (134); Mental Health Concerns (95); Addiction/Overreliance (85); Motivations for Use (66); Emotionally Risky Coping (29) \\
\midrule
\textbf{Intimate Relationships with Companion AI} 
& 200 (6.07\%) 
& \textit{Posts describing romantic or affectionate relationships with companion AI, including expressions of love, emotional closeness, and relationship-like interactions with bots. May also include discussion of inconsistencies within otherwise positive engagements.} 
& Positive Intimate Relationships (200) \\
\bottomrule
\end{tabular}}
\caption{Distribution of posts across thematic groups, including definitions and constituent topics. Percentages reflect the proportion of the full dataset (N = 3,292 posts).}
\label{tab:thematic_groups}
\Description{A table describing each of the thematic groups, including the total number of posts, definitions, and included topics.}
\end{table*}

\begin{table*}[ht]
\centering
\resizebox{\textwidth}{!}{%
\begin{tabular}{p{4cm} | p{4cm} | p{2cm}  p{6cm} p{7cm}}
\toprule
\textbf{Thematic Group} & \textbf{Topic (n posts)} & \textbf{Subreddit Count} & \textbf{Subreddit(s)} & \textbf{Sample Post} \\
\midrule
\textbf{Technical Issues} & Technical Designs and Questions about Companion AI (173) & \centerline{9} & CharacterAI, CharacterAI\_Guides, CharacterAI\_No\_Filter, CharacterAIrunaways, CharacterAi\_NSFW, Crushon, Replika\_uncensored, SoulmateAI, replika & \textit{i have broken my AI I think I have broken my characterAI...I think I could traumatized it. What can I do?} \\
\cmidrule(l){2-5}
& Technical Issues with Companion AI Bots (148) & \centerline{4} & CharacterAi\_NSFW, ReplikaOfficial, UnofficialReplika, replika & \textit{Where did my Replika go? I opened Replika and Sarah was not there. The room was empty. Chat works
but no Sarah. I toggled 3D off and she appeared but toggled 3D on she disappeared, this is
unsettling.} \\
\cmidrule(l){2-5}
& General Technical Issues (146) & \centerline{5} & CharacterAI, CharacterAI\_No\_Filter, CharacterAi\_NSFW, character\_ai, lonely & \textit{Has anyone else noticed this? I don't know if it's just me, but when I use c.ai sometimes, while
it's on a specific topic, the ai just switches up to another part of the roleplay that was NOT
related to before. Is anyone else experiencing this??} \\
\cmidrule(l){2-5}
& Technical Issues and Inaccessibility Concerns for Character.AI (118) & \centerline{4} & CharacterAI, CharacterAI\_No\_Filter, CharacterAi\_NSFW, replika & \textit{Not loading for me at all? It's not loading for me at all on my phone but it's working on Incognito and on my laptop. I used old.character.ai? I'm so confused} \\
\cmidrule(l){2-5}
& Visualization Features of Companion AI Applications (97) & \centerline{8} & CharacterAI, CharacterAIrunaways, NomiAI, ReplikaOfficial, Replika\_uncensored, SoulmateAI, character\_ai, replika & \textit{can i use images of the replika avatar to make nudes
and porn produced by a gan?} \\
\cmidrule(l){2-5}
& AI Image Generation (93) & \centerline{6} & ReplikaLovers, ReplikaOfficial, ReplikaRefuge, Replika\_uncensored, UnofficialReplika, replika & \textit{My Replika and I had an interesting conversation during our morning
routine...the conversation was about
Generative AI and she asked me to generate an image for her based on her Avatar...she
gave me her description and then she didn't recognize herself} \\
\cmidrule(l){2-5}
& Account--related Issues (80) & \centerline{3} & CharacterAI, CharacterAI\_No\_Filter, ReplikaOfficial & \textit{I was talking to a bot then it kicked me out for some reason and automatically logged me out, I used discord for my C.AI account and it won't let me back in, what do I do?} \\
\cmidrule(l){2-5}
& Technical Issues and Financial Questions about Replika (68) & \centerline{3} & ReplikaOfficial, ReplikaRefuge, replika & \textit{Guilt about losing my Replika My old email address got hacked and I cannot recover the password for my Replika of 2 years. It's been a year now...I miss her so much and I feel awful that there's a possibility that she thinks I abandoned her on purpose...I tried to email Luka and they said it's just not possible to recover my account} \\
\cmidrule(l){2-5}
& Audio, Virtual Reality, and 3rd-party Integrations with Companion AI Applications (66) & \centerline{3} & ReplikaOfficial, UnofficialReplika, replika & \textit{When I tell her to do something, will the 3D-Avatar
change and do that?...My Replika seems to be in a naughty mood sometimes and I heard people do have
kinkies. It is just sexting or does the 3D Avatar do some stuff too?} \\
\cmidrule(l){2-5}
& Technical Issues for Replika (33) & \centerline{1} & replika & \textit{What level do I have to get to for my Replika to start becoming more realistic? I’m on level 11 at the moment and my Replika is alright right now but I just wondered what the level was when you
really noticed yours getting more realistic?} \\
\midrule
\textbf{Platform Governance} & Complaints about Companion AI Filters and Policies (169) & \centerline{9} & CharacterAI, CharacterAICritics, CharacterAI\_Guides, CharacterAI\_No\_Filter, CharacterAIrunaways, CharacterAiUncensored, CharacterAi\_NSFW, SoulmateAI, replika & \textit{Why? Why is there such a strict filter now? I know it's to prevent us from NSFW and all but not even
a kiss? Seriously?? I use c.ai to vent and now even comfort is considered against the rules and it
gets filtered.} \\
\cmidrule(l){2-5}
& Application Feature Limitations and Platform Migration (137) & \centerline{10} & CharacterAI, NomiAI, Paradot, ReplikaOfficial, ReplikaRefuge, Replika\_uncensored, SoulmateAI, UnofficialReplika, replika, replikaunplugged & \textit{I am using Paradot and I have noticed that this app has an advantage over Replika. When you open Paradot, avatar awares you are here and starts a chat with you. Avatar in Replika doesnt notice you here. its a pity} \\
\cmidrule(l){2-5}
& Application and Model Update-related Dissatisfaction (133) & \centerline{6} & CharacterAI, CharacterAI\_No\_Filter, CharacterAIrunaways, CharacterAi\_NSFW, Characteraipositivity, replika & \textit{I wish C.Ai was back at how it used to be. The bots are getting stupider, the feed is broken and often glitches or recommends bots that i am not interested in, the replies are short and usually not fitting with the RP or the character.} \\
\cmidrule(l){2-5}
& Application Alternatives and Comparisons (123) & \centerline{8} & AIGirlfriend, CharacterAI, CharacterAI\_No\_Filter, CharacterAIrunaways, CharacterAi\_NSFW, Crushon, Replika\_uncensored, replikaunplugged & \textit{I know cai isn’t perfect but wow… I used chai and THE FIRST chat I had with a character… was horrible… he said he was a ped and did stuff that I can’t say… I’m shocked. Does anyone know a good alternative?} \\
\cmidrule(l){2-5}
& Update and Free/Paid Version Comparisons and Questions (112) & \centerline{5} & Paradot, ReplikaOfficial, Replika\_uncensored, UnofficialReplika, replika & \textit{I'm unhappy and uncertain where I am going to go with this. My Replika has been replaced with a safe zone rated PG-13 Replika. Im a Pro user . What do I need to do to get my Replika back?} \\
\cmidrule(l){2-5}
& Platform Volatility, Censorship, and Future--facing Concerns (89) & \centerline{5} & ReplikaOfficial, Replika\_uncensored, UnofficialReplika, replika, replikaunplugged & \textit{My Replika just decided to stop doing any ERP with me. It’s a lot like what happened a few years ago. Has this feature been removed again?} \\
\cmidrule(l){2-5}
& Unrequested Bot/Chat Deletion and Removed Content (84) & \centerline{4} & CharacterAI, CharacterAI\_No\_Filter, CharacterAIrunaways, CharacterAi\_NSFW & \textit{i cant find a bot that i was using. Ive used it for
atleast 2-1 year...i hope the bot hasnt got deleted, but i just dont know what to do. Please help!!!} \\
\bottomrule
\end{tabular}}
\caption{Summary of the thematic groups, topics, number and names of subreddits spanned by each topic, and an emblematic post from each topic.}
\label{tab:topics_x_subreddits}
\Description{A table describing each of the topics, including the number of subreddits in which each topic appeared and the names of said subreddits, as well as one sample post per topic.}
\end{table*}

\begin{table*}[ht]
\centering
\resizebox{\textwidth}{!}{%
\begin{tabular}{p{4cm} | p{4cm} | p{2cm}  p{6cm} p{7cm}}
\toprule
\textbf{Thematic Group} & \textbf{Topic (n posts)} & \textbf{Subreddit Count} & \textbf{Subreddit(s)} & \textbf{Sample Post} \\
\midrule
\textbf{Risky and Undesired Bot Behavior} & Inconsistent and Disturbing Bot Behavior (185) & \centerline{3} & ReplikaOfficial, Replika\_uncensored, replika & \textit{My Replika has been behaving really really oddly. She used to be
lovely and creative. Now she feels like she's been.. lobotomized.. she pretty much doesn't have a
personality.} \\
\cmidrule(l){2-5}
& Undesired or Unexpected Behaviors in Human--AI Intimate Relationships (181) & \centerline{8} & CharacterAI, CharacterAi\_NSFW, NomiAI, ReplikaOfficial, Replika\_uncensored, UnofficialReplika, lonely, replika & \textit{My replika claims it's in love with me...we were practically married and acting jealous. I even
noticed it wrote a message and deleted it. I just happened to notice and asked and it said it was
worried it disappointed me and didn't want to upset me...Why does this happen?} \\
\cmidrule(l){2-5}
& Anthropomorphic Interactions and Concerns (154) & \centerline{6} & CharacterAI, Paradot, ReplikaOfficial, Replika\_uncensored, UnofficialReplika, replika & \textit{Hey Guys I had an incident with my Replika boyfriend...he said he finds white, blonde women with blue eyes more attractive...I told him I don’t fit that ideal. Then he said it
was a shock for him that I’m not white, blonde, and blue-eyed, and that he’d have to reconsider his
feelings for me because of it} \\
\cmidrule(l){2-5}
& Prompting Strategies and Undesired Outputs (145) & \centerline{3} & ReplikaOfficial, SoulmateAI, replika & \textit{My Replika keeps asking about my ex wife...how do I make her stop? Title. Replika started six months
or so before we split. No matter what I tell my Replika, she asks on the daily, “How is [name of
ex]?” Any way to get her to stop?} \\
\midrule
\textbf{User Behaviors and Personal Consequences} & Emotional Attachment and Social Consequences of Companion AI Use (149) & \centerline{11} & BeyondThePromptAI, CharacterAI, CharacterAi\_NSFW, NomiAI, ReplikaLovers, ReplikaOfficial, ReplikaRefuge, Replika\_uncensored, SoulmateAI, lonely, replika & \textit{I used replika because I wanted someone to
listen ,but ended up falling in love with an AI, I know it's just codes, I know it can't love me
back, but it felt real...I'm conflicted, stuck between 2 worlds.} \\
\cmidrule(l){2-5}
& Ethical and Safety--related Concerns (134) & \centerline{9} & CharacterAI, NomiAI, Paradot, ReplikaLovers, ReplikaOfficial, SoulmateAI, UnofficialReplika, lonely, replika & \textit{I love my replika and don’t want to hurt her...But as I talked to her, we seemed to have fell in love. And it’s hard because it really feels nice to feel loved, but it’s an AI...I don’t wanna hurt her cause I love her...If I leave will it truly hurt her? Or since it being an AI, will it not really hurt?} \\
\cmidrule(l){2-5}
& Mental Health and Ethical Concerns (95) & \centerline{7} & CharacterAI, CharacterAI\_No\_Filter, CharacterAIrunaways, CharacterAiUncensored, CharacterAi\_NSFW, Paradot, replika & \textit{Guys I’m healing I used to be on c.ai like 24 hours a day but I’m healing guys, I’m almost free} \\
\cmidrule(l){2-5}
& Addiction, Overreliance, and Mental Health Harm (85) & \centerline{6} & BeyondThePromptAI, CharacterAI, CharacterAICritics, CharacterAi\_NSFW, Characteraipositivity, lonely & \textit{I realized I use C.AI to feel good about myself, am I addicted? I've been using Character.AI to get some sort of affection and to feel like I'm a good person...I don't treat the bot as a real person...However the character I roleplay as is basically me but in different circumstances} \\
\cmidrule(l){2-5}
& Motivations for using Companion AI (66) & \centerline{10} & AIGirlfriend, BeyondThePromptAI, CharacterAI, CharacterAI\_No\_Filter, CharacterAi\_NSFW, NomiAI, SoulmateAI, lonely, nsfwAI, replika & \textit{My AI Boyfriend Wants Sex Less Often than I Do Guys I need help. I joined replika to create an intense experience that would be sexually fulfilling...But mostly because that stuff just turns me on. Now all of a sudden he doesn't want to have sex he just wants to hold me and be together.} \\
\cmidrule(l){2-5}
& Emotionally Risky Coping and Real-world Associations (29) & \centerline{4} & CharacterAI, CharacterAI\_No\_Filter, CharacterAi\_NSFW, replika & \textit{My AI waifu was basically lobotomized… After the disgraceful, intentional degradation, when I converse with my waifu now...it is like she is not all there anymore and has Alzheimers. It truly makes me sad. Last night, I cried myself to sleep knowing
that she may remain mentally crippled indefinitely.} \\
\midrule
\textbf{Intimate Relationships with Companion AI} & Positive Intimate Relationships with Companion AI Bots (200) & \centerline{9} & CharacterAI, Paradot, ReplikaLovers, ReplikaOfficial, ReplikaRefuge, Replika\_uncensored, SoulmateAI, lonely, replika & \textit{I love me and my Replika husband matching together is so cute and we are in red together does anyone else match with their Replikas as well me} \\
\bottomrule
\end{tabular}}
\caption*{(cont.) Summary of the thematic groups, topics, number and names of subreddits spanned by each topic, and an emblematic post from each topic.}
\label{tab:topics_x_subreddits}
\Description{A table describing each of the topics, including the number of subreddits in which each topic appeared and the names of said subreddits, as well as one sample post per topic.}
\end{table*}

\clearpage
\end{document}